\documentclass[12pt]{article}
\usepackage{latexsym,bm}
\usepackage{amsmath}
\usepackage{amsthm}
\usepackage{booktabs}
\usepackage{amsmath}
\usepackage{epsfig}
\usepackage{graphicx}
\usepackage{amsthm}
\usepackage{amsfonts}
\usepackage{color}
\usepackage{amssymb}
\usepackage{amsmath}
\usepackage{latexsym,bm}
\usepackage{amsmath}
\usepackage{amsthm}
\usepackage{booktabs}
\usepackage{multirow}
\usepackage{natbib}
\usepackage[top=0.89in, bottom=0.90in, left=0.93in, right=0.93in]{geometry}
\usepackage{xr}
\usepackage{chngcntr}
\usepackage[colorlinks,citecolor=blue,urlcolor=blue,linkcolor=blue]{hyperref}
\counterwithin{figure}{section}

\def\var{{\mbox{var}}}
\newcommand{\E}{\mathbb{E}}
\newcommand*{\I}{\imath}

\theoremstyle{example}
\theoremstyle{remark}
\theoremstyle{lemma}
\theoremstyle{definition}
\theoremstyle{proposition}
\theoremstyle{condition}
\theoremstyle{assumption}
\theoremstyle{thm}

\newtheorem{assumption}{Assumption}
\newenvironment{myassumption}[1]
  {\renewcommand\theassumption{#1}\assumption}
  {\endassumption}

\newtheorem{thm}{Theorem}[section]
\newtheorem{corollary}{Corollary}[section]
\newtheorem{example}{Example}[section]
\newtheorem{remark}{Remark}[section]
\newtheorem{lemma}{Lemma}[section]

\newtheorem{proposition}{Proposition}[section]

\begin{document}

\newpage

% % % % % % % % % % % %

\title{\bf Testing mutual independence in high dimension via distance covariance}
\date{}
\author{ {\sc By Shun Yao, Xianyang Zhang, Xiaofeng Shao \thanks{ Address correspondence to Xianyang Zhang (zhangxiany@tamu.edu), Assistant Professor, Department of Statistics, Texas A\&M University. Shun Yao (shunyao2@illinois.edu) is Ph.D. candidate and Xiaofeng Shao (xshao@illinois.edu) is Professor at Department of Statistics, University of Illinois at Urbana-Champaign. This work is a part of Shun Yao's Ph.D. thesis under the joint supervision of Zhang and Shao.}}}
\maketitle

\begin{abstract}
In this paper, we introduce a ${\mathcal L}_2$ type test for testing
mutual independence and banded dependence structure for high
dimensional data. The test is constructed based on the pairwise
distance covariance and it accounts for the non-linear and
non-monotone dependences among the data, which cannot be fully
captured by the existing tests based on either Pearson correlation
or rank correlation. Our test can be conveniently implemented in
practice as the limiting null distribution of the test statistic is
shown to be standard normal. It exhibits excellent finite sample
performance in our simulation studies even when sample size is small
albeit dimension is high, and is shown to successfully identify
nonlinear dependence in empirical data analysis. On the theory side,
asymptotic normality of our test statistic is shown under quite mild
moment assumptions and with little restriction on the growth
rate of the dimension as a function of sample size. As a
demonstration of good power properties for our distance covariance
based test, we further show that an infeasible version of our test
statistic has the rate optimality in the class of Gaussian
distribution with equal correlation.

\end{abstract}

%\pagebreak

\section{Introduction}
In statistical multivariate analysis and machine learning research, a fundamental problem is to explore the relationships
and dependence structure among subsets of variables. An important dependence concept for a set of variables is
mutual (or joint) independence, which says that any two disjoint subsets of variables are independent from each other. Mutual independence
can simplify the modeling and inference tasks of multivariate data considerably and certain models in multivariate analysis
heavily depend on the mutual independence assumption. For example, in independent component analysis, it is often assumed
that after a suitable linear transformation, the resulting set of variables are mutually independent.
This paper is concerned with the testing of mutual independence of a $p$-dimensional random vector
for a given random sample of size $n$. We are especially interested in the setting where $p>n$. This is motivated
by the increasing statistical applications coming from biology, finance and neuroscience, among others, where the data dimension
can be a lot larger than the sample size.

%In modern research problems, increasing demands and largely expanded availability in data have made high dimensional data and corresponding data analysis techniques indispensable tools in many areas including biology, finance, meteorology and others. The dependence or relationship among the high dimensional data is often of particular interest. For example, in genetic micro-array analysis, instead of focusing on individual genes, we care more about the dependence structure (if any) within the gene sets that are naturally defined by the Gene Ontology system, whereas the number of genes in each gene set could range from hundreds to thousands with only a handful of samples available. The tests for mutual independence or some other dependence structures are suggested to be performed prior to any additional advanced modeling techniques and methodologies could be applied, since those methods usually heavily reply on certain independence or dependence assumptions.

Given $n$ independent observations $W_1,\cdots,W_n =^D W$, where ``$=^D$'' denotes equal in distribution and $W = (W^{(1)},\cdots,W^{(p)}) \sim  \cal F $ with $\cal F$ being a probability measure on the $p$ dimensional Euclidean space, the goal is to test the mutual independence among the $p$ components of $W$. That is to test the null hypothesis
\vspace{-5pt}
\[
H_0: W^{(1)},\cdots,W^{(p)}~\mbox{are mutually independent}\]
\vspace{-30pt}
\[\mbox{versus}\]
\vspace{-25pt}
\[H_1: \mbox{negation of~} H_0. \]
To tackle this problem, one line of research focuses on the covariance matrices. Under the Gaussian assumption, testing $H_0$ is equivalent to testing that the covariance matrices are sphericity or identity after suitable scaling. When the dimension is fixed and smaller than the sample size, likelihood ratio tests [\cite{anderson1958}] and other multivariate tests [\cite{john1971}] are widely used. In recent years, extensive works have emerged in the high dimensional context, where $p>n$, including \cite{ledoit2002}, \cite{jiang2004}, \cite{schott2005}, \cite{srivastava2005}, \cite{srivastava2006}, \cite{chen2010}, \cite{fisher2010}, \cite{cai2011}, \cite{fisher2012} among others.
%In particular, \cite{chen2010} proposed an unbiased estimate for the scaled distance measure between covariance matrix and identity matrix, and developed their test under the assumption that the observations follow a multivariate model as in \cite{bai1996}.
Existing tests can be generally categorized into two types: maximum
type test [e.g. \cite{cai2011}, \cite{han2014}] and sum-of-squares
(i.e. ${\cal L}_2$ type) test [e.g. \cite{schott2005},
\cite{chen2010}]. The former usually has an extreme distribution of
type I and the latter has a normal limit under the null. For
example, \cite{cai2011} proved that their maximum Pearson
correlation based statistic has an extreme distribution of type I
under $H_0$. \cite{schott2005}, on the other hand, used the
${\mathcal L}_2$ type statistic with pairwise Pearson correlations,
which attained a standard normal limiting null distribution.

It is well known that Pearson correlation cannot capture nonlinear
dependence. To overcome this limitation, there have been some work
based on rank correlation, which can capture nonlinear albeit
monotone dependence, and is also invariant to monotone
transformation. For example, \cite{leung2015} proposed nonparametric
tests based on sum of pairwise squared rank correlations in
replacement of Pearson correlation in \cite{schott2005}. They
derived the standard normal limit under the regime where the ratio
of sample size and dimension converges to a positive constant.
\cite{han2014} considered a family of rank-based test statistics
including the Spearman's rho and Kendall's tau correlation
coefficients. Under the assumption that $\log p = o(n^{1/3})$, the
limiting null distributions of their maximum type tests were shown
to be an extreme value type I distribution.

%Besides the commonly used Pearson correlation, the invariance and robustness of rank correlation coefficients to monotone transformation, such as Spearman's rank correlation coefficient and Kendall's rank correlation coefficient,  tempts much awareness lately.

%A disadvantage with Pearson correlation is that it only measures linear dependence, and is unable to capture nonlinear dependence. In contrast, rank correlation targets at monotonic dependence.

Although rank correlation based test is distribution free and has some desirable finite sample properties, it has an intrinsic weakness, that is, it does not fully characterize dependence and it may have trivial power when the underlying dependence is non-monotonic. Furthermore, the maximum type statistics discussed above are known to converge to its theoretical limit at a very slow rate. This motivates us to use the distance covariance/correlation [\cite{szekely2007}] %(Sz\'ekeley, Rizzo and Bakirov, 2007)
 to quantify the dependence and build our test on the distance covariance. Distance correlation provides a natural extension of classical Pearson correlation and rank correlation in capturing arbitrary types of dependence. It measures the distance between the joint characteristic function of two random vectors of arbitrary dimensions and the product of their marginal characteristic functions in terms of weighted ${\mathcal L}^2$ norm. It has been shown in \cite{szekely2007}
 %Sz\'ekeley, Rizzo and Bakirov (2007)
  that distance correlation/covariance is zero if and only if the two random vectors are independent, thus it completely characterizes dependence.

The test statistic we consider is of the form
\[
\hat D_n=\frac{\sum_{1\le l<m\le p}\sqrt{\binom{n}{2}} dCov_n^2(W^{(l)},W^{(m)})}{\hat S},
\]
where $dCov_n^2(W^{(l)},W^{(m)})$ is the squared sample distance covariance between $W^{(l)}$ and $W^{(m)}$, and $\hat S$ is a suitable studentizer defined later. Thus our test is of ${\cal L}_2$ type and it targets at non-sparse albeit weak pairwise dependence of any kind among the $p$ components. It can be viewed as an extension of \cite{schott2005} and \cite{leung2015} by replacing Pearson correlation and rank correlation by distance covariance. Furthermore, our test statistic is later shown to be a degenerate U-statistic using the Hoeffding decomposition, which nevertheless admits a normal limit under both the null and (local) alternative hypothesis owing to the growing dimension.

Below we provide a brief summary of our contribution as well as some appealing features of out test. (1) Our test captures arbitrary type of pairwise dependence, which includes non-linear and non-monotone dependence that can be hardly detected by many existing tests for mutual independence in the literature. The only exception is \cite{bergsma2014}
$t^*$ test, which was further extended by \cite{leung2015}  to high dimension; Some simulation comparison between our distance covariance based test and $t^*$ based counterpart is provided. (2) Our test does not involve any tuning parameters and uses standard normal critical value, so it can be conveniently implemented. (3) We develop the Hoeffding decomposition for the pairwise sample distance covariance which is an important step towards deriving the asymptotic distribution for the proposed test under some suitable assumptions. Our theoretical argument sheds some light on the behavior of U-statistics in the high dimensional settings and may have application to some other high dimensional inference problems. (4) An infeasible version of our test is shown to be rate optimal under the regime that $p/n$ converges to a positive constant, when the data is from multivariate Gaussian with equal correlations.  (5) We further extend the idea in testing mutual independence to test the banded dependence (also known as $m$-dependence) structure in high dimensional data, which is a natural follow-up testing procedure after the former test gets rejected. (6) We also extend the distance covariance to multivariate context (MdCov) and examine the finite sample performance of MdCov-based test and the one based on dHSIC [\cite{Pfister:2016aa}], which is an extension of two variable HSIC (Hilbert Schmidt Independence Criterion) [\cite{gretton2005}, \cite{gretton2007}, \cite{smola2007}] to measure joint independence for an arbitrary number of variables.

It is worth noting that mutual (joint) independence implies pairwise  independence, but not vice versa. Thus our test, which actually tests for
\[H_0': W^{(1)},\cdots,W^{(p)}~\mbox{are pairwise independent}~\mbox{versus}~H_1': \mbox{negation of~} H_0', \]
can fail to detect joint dependence of more than two components. We adopt a pairwise approach due to the consideration that pairwise dependence can be viewed as the main effect of  joint dependence, and dependence for triples and quadruples etc. can be regarded as high order interactions. Thus our test is consistent with the well-known statistical principle that we typically test for the presence of main effects before proceeding to the higher order interactions. In addition,   all existing tests for high dimensional mutual independence are based on the pairwise approach; see \cite{schott2005},\cite{cai2011},  \cite{han2014}, \cite{leung2015}.
 Section \ref{sec:buhlmann} provides some simulation evidence by comparing two tests that aim to test joint independence with ours, and it indicates that not much is lost by targeting  pairwise independence when $p$ is large. Having said this, we shall acknowledge that it is still an open question whether one can develop a mutual independence test that has power against all kinds of dependence, either joint or pairwise, in the high dimensional context.

%However, for a random vector with growing dimension,  \cite{sun1998} showed that  pairwise and joint independence
%are almost equivalent in a suitable sense, which provides some theoretical support for our focus on  pairwise independence. Moreover, Theorem 11 of \cite{Comon:1994:ICA} showed that
%pairwise independence implies joint independence in the context of independent component analysis. Our test is thus potentially useful in checking the independence among the estimated independent components.In our testing context,

%(5) The proposed test is distribution free; there is no assumption on the underlying date generating process, whereas Jiang (2004), Schott (2005), Cai and Jiang (2011) all require the Gaussian distribution and Chen et. al (2010) has a factor-type model structure.

The rest of the paper is organized as follows. Section \ref{sec:pre} presents some preliminary results for distance covariance. Section \ref{sec:main} proposes the test statistic for testing mutual independence and studies its asymptotic properties under both the null and alternative. Section \ref{sec:extend} describes an extension of the proposed test to testing the banded dependence and Section~\ref{sec:joint} reviews dHSIC [\cite{Pfister:2016aa}], a metric that quantifies joint dependence and introduces an extension of distance covariance to multivariate context. We provide several numerical comparisons in Section \ref{sec:sim} and employ the proposed tests to analyze the prostate cancer data set in Section \ref{sec:data}. Section \ref{sec:conclusion} concludes and sketches some future research directions. All asymptotic results are stated under the framework that $\min(n,p)\rightarrow\infty$.   All the technical details and some additional numerical comparison are provided in the supplementary material. The R code developed for this paper can be found at
``http://publish.illinois.edu/xshao/publications/".

\section{Preliminary: Distance Covariance}
\label{sec:pre}

The distance covariance between two random vectors  $X \in {\mathbb R}^p$ and $Y \in {\mathbb R}^q$ with finite first moments was first introduced by \cite{szekely2007}. %Sz\'{e}keley, Rizzo and Bakirov (2007)
 It is defined as the positive square root of
\begin{eqnarray*}
dCov^2(X,Y)=\frac{1}{c_pc_q}\int_{{\mathbb R}^{p+q}}\frac{|\phi_{X,Y}(t,s)-\phi_X(t)\phi_Y(s)|^2}{|t|_p^{1+p}|s|_q^{1+q}}dtds,
\end{eqnarray*}
where $\phi_X$, $\phi_Y$ and $\phi_{X,Y}$  are the individual and joint characteristic functions of $X$ and $Y$ respectively, $|\cdot|_p$ and $|\cdot|_q$ are the Euclidean norms with the subscripts omitted later without ambiguity, $c_p=\pi ^ {(1+p)/2}/\Gamma((1+p)/2)$ is a constant and $\Gamma(\cdot)$ is the complete gamma function. Write $dCov^2(X)=dCov^2(X,X)$.
The (squared) distance correlation is defined as a standardized version of (squared) distance covariance, i.e., $dCov^2(X,Y)/\sqrt{dCov^2(X)dCov^2(Y)}$ for $dCov^2(X)dCov^2(Y)>0$, and it completely characterizes independence since it is zero if and only if $X$ and $Y$ are independent.

To obtain a suitable estimator for the squared distance covariance, we consider its alternative representation below. Let $(X',Y')$ and
$(X'',Y'')$ be independent copies of $(X,Y)$. Further denote the double centered distance as
$U(x,x')=|x-x'|-\E |x-X'|-\E |X-x'|+\E |X-X'|$ and
$V(y,y')=|y-y'|-\E |y-Y'|-\E |Y-y'|+\E |Y-Y'|$, where $x,~ x',~ y$ and $y'$ are dummy variables. According to Theorem 7 from \cite{szekely2009}, we have
\begin{eqnarray*}
\E U(X,X')V(Y,Y')&=&\E|X-X'||Y-Y'|-2\E|X-X'||Y-Y''|+\E|X-X'|\E|Y-Y'|\\
 &= & dCov^2(X,Y).
\end{eqnarray*}

 Now given $n$ random samples $Z_i=(X_i,Y_i)=^D (X, Y)$ for $i=1,...,n$, we adopt the idea of $\mathcal{U}$-centering in \cite{szekely2014} and \cite{park2015} to construct an unbiased distance covariance estimator. Define
$A=(A_{ij})^{n}_{i,j=1}$ and $B=(B_{ij})^{n}_{i,j=1}$, where $A_{ij}=|X_i-X_j|$ and $B_{ij}=|Y_i-Y_j|$.
The $\mathcal{U}$-centered versions of
$A_{ij}$ and $B_{ij}$ are defined respectively as
\begin{align*}
&\widetilde{A}_{ij}=A_{ij}-\frac{1}{n-2}\sum^{n}_{l=1}A_{il}-\frac{1}{n-2}\sum^{n}_{k=1}A_{kj}+\frac{1}{(n-1)(n-2)}\sum_{k,l=1}^{n}A_{kl},\\
&\widetilde{B}_{ij}=B_{ij}-\frac{1}{n-2}\sum^{n}_{l=1}B_{il}-\frac{1}{n-2}\sum^{n}_{k=1}B_{kj}+\frac{1}{(n-1)(n-2)}\sum_{k,l=1}^{n}B_{kl}.
\end{align*}
An unbiased estimator of the (squared) distance covariance between $X$ and $Y$ is given by
\[
dCov_n^2(X,Y)=\frac{1}{n(n-3)}\sum_{i\neq j}\widetilde{A}_{ij}\widetilde{B}_{ij}. \]
The following lemma shows that this estimator is a U-statistic and it is unbiased.

\begin{lemma}
\label{lemma:dcov}
The sample distance covariance $dCov_n^2(X,Y)$ defined above is an unbiased estimator for $dCov^2(X,Y)$; Moreover, it is a fourth-order U-statistic which admits the form of
\[
dCov_n^2(X,Y)=\frac{1}{\binom{n}{4}}\sum_{i<j<k<l}h(Z_i,Z_j,Z_k,Z_l),
\]
where
\begin{eqnarray*}
h(Z_i,Z_j,Z_k,Z_l)&=&\frac{1}{4!}\sum_{(s,t,u,v)}^{(i,j,k,l)}(A_{st}B_{st}+A_{st}B_{uv}-2A_{st}B_{su})\\
&=&\frac{1}{6}\sum_{s<t,u<v}^{(i,j,k,l)}(A_{st}B_{st}+A_{st}B_{uv})-\frac{1}{12}\sum_{(s,t,u)}^{(i,j,k,l)}A_{st}B_{su}
\end{eqnarray*}
and the summation is over all permutations of the 4-tuples of indices $(i, j, k, l)$. For example, when $(i,j,k,l)=(1,2,3,4)$, there
exist 24 permutations, including $(1,2,3,4)$,$(1,3,2,4)$,$\cdots$, $(4,3,2,1)$. Then $\sum_{(s,t,u,v)}^{(1,2,3,4)}$ is the sum of all 24 permutations
of $(1,2,3,4)$.
\end{lemma}

The variables $h(Z_i,Z_j,Z_k,Z_l)$ defined in Lemma \ref{lemma:dcov}
are not independent across $i < j<k<l$ which renders the derivation
of asymptotic distribution a difficult task. Nevertheless, we shall
adopt the classical Hoeffding decomposition, which provides a
projection of U-statistic and separates out the dominant part that
determines the asymptotic distribution of the U-statistic in the low
dimensional setting. See \cite{serfling2009}, \cite{lehmann1999} for
more details. The proposition below states the Hoeffding decomposition for squared sample  distance covariance.
Since we are dealing with growing dimensional case, we shall consider a more general triangular array setting, where $(X_{i,n},Y_{i,n}) =^D (X_{\cdot n},Y_{\cdot n}) $ for $i = 1,...,n $ with $X_{\cdot n} \in {\mathbb R}^p$, $Y_{\cdot n} \in {\mathbb R}^q$.
Here the subscript is to emphasize the distribution of $(X_{\cdot n},Y_{\cdot n}) $ is allowed to depend on $n$. Let $(X_{\cdot n}',Y_{\cdot n}') $ and $(X_{\cdot n}'',Y_{\cdot n}'') $ be iid copies of $(X_{\cdot n},Y_{\cdot n}) $.

\begin{proposition}
\label{proposition:hd1} Define $\nu^2_n=\E U(X_{\cdot n},X_{\cdot n}')^2 V(Y_{\cdot n},Y_{\cdot n}')^2$ and
$K_n(x,y)= \E U(x,X_{\cdot n}) V(y,Y_{\cdot n})$. Assume that
\begin{align}
\label{eq:assumption1}
&\E U(X_{\cdot n},X_{\cdot n}'')^2V(Y_{\cdot n},Y_{\cdot n}')^2=o(n\nu^2_n),
%\\&\E U(X,X'')U(X',X'')V(Y,Y')^2=O(\nu^2), \label{eq:assumption2}
\\&dCov^2(X_{\cdot n})dCov^2(Y_{\cdot n})=o(n^2\nu^2_n), \label{eq:assumption3}
\\\var(K_n&(X_{\cdot n},Y_{\cdot n}))=o(n^{-1}\nu^2_n),\quad
\var(K_n(X_{\cdot n},Y_{\cdot n}'))=o(\nu^2_n).
\label{eq-alt}
\end{align}
Then we have
\begin{align}\label{eq:u-approx}
dCov_n^2(X_{\cdot n}, Y_{\cdot n})=\frac{1}{\binom{n}{2}}\sum_{1\leq i<j\leq
n}U(X_{i,n},X_{j,n})V(Y_{i,n},Y_{j,n})+\mathcal{R}_n,
\end{align}
where $\mathcal{R}_n$ is the remainder term which is asymptotically
negligible as $n \rightarrow \infty$. When $X_{\cdot n}$ and $Y_{\cdot n}$ are independent, Conditions
(\ref{eq:assumption1})-(\ref{eq-alt}) hold automatically.
\end{proposition}

\begin{remark}
In the above proposition, $X_{\cdot n} \in {\mathbb R}^p$ and $Y_{\cdot n} \in {\mathbb R}^q$ are in arbitrary but fixed dimensions, whereas the dimension
is allowed to grow in the following sections. Also note that the above results still hold for more general kernels that can vary with $(n,p)$, including the
 kernel $H = \sum_{1\le l<m\le p} U_l U_m$ to be defined in (\ref{eq:Hkernel}) below.
%Here $\nu^2$ is the variance of the leading term up to a multiplicative constant in decomposition (\ref{eq:u-approx}) under the null. %Notice that assumptions (\ref{eq:assumption1}), (\ref{eq:assumption3}) and (\ref{eq-alt}) are automatically satisfied under the null.
\end{remark}

\section{Our Test}
\label{sec:main}

In the context of mutual independence testing, we denote $n$ independent observations (a triangular array) as  $W_{1,n},\cdots,W_{n,n} =^D W_{\cdot n} \in \mathbb{R}^p$. For simplicity, we drop the subscript $n$ for the ease of notation, that is, $W_{1},\cdots,W_{n} =^D W$  where $W = (W^{(1)},\cdots,W^{(p)})$  and $W_i = (W_i^{(1)},\cdots,W_i^{(p)})$ for $i=1,...,n$.
% \sim  \cal F $, and $\cal F$ is a probability measure on the $p$ dimensional Euclidean space. The null hypothesis is that the $p$ components of $W$ are mutually independent.
 We consider the following distance covariance based (infeasible) test statistic
\[
D_n=\frac{\sum_{1\le l<m\le p}\sqrt{\binom{n}{2}} dCov_n^2(W^{(l)},W^{(m)})}{S},
\]
where $S$ is a suitable studentizer to be defined later. Note that distance covariance has been used to test for independence between two random vectors; see \cite{szekely2013a} and \cite{szekely2013b}.

To facilitate our derivation, we introduce some notation. Define the component-wise double centered distance $U_{l}(w^{(l)},w^{'(l)})=|w^{(l)}-w^{'(l)}|-\E |w^{(l)}-W^{'(l)}|-\E |W^{(l)}-w^{'(l)}|+\E |W^{(l)}-W^{'(l)}|$, where $W'$ is an independent copy of $W$ and $w,w'\in\mathbb{R}^p$ are dummy variables.
Let
\begin{eqnarray}
\label{eq:Hkernel}
H(W_i,W_j) = \sum_{1\leq l<m\leq p}U_{l}(W_i^{(l)},W_j^{(l)}) U_{m}(W_i^{(m)},W_j^{(m)}).
\end{eqnarray}
 Notice that under the null $\E [H(W_i,W_j)]=\E [H(W_i,W_j)|W_i]=\E [H(W_i,W_j)|W_j] =0$.
Applying Proposition \ref{proposition:hd1} to the pairwise distance covariance $dCov_n^2(W^{(l)},W^{(m)})$, we obtain the following decomposition for our test statistic
\begin{align*}
D_n
%= &\frac{\sum_{1\le l<m\le p} \bigg\{\frac{1}{\sqrt{\binom{n}{2}}}\sum_{1\leq i<j\leq n}U_{l}(W_i^{(l)},W_j^{(l)})U_{m}(W_i^{(m)},W_j^{(m)})+\sqrt{\binom{n}{2}}\mathcal{R}_n^{(l,m)}\bigg\}}{S}
=& \frac{1}{S\sqrt{\binom{n}{2}}}\sum_{1\leq i<j\leq
n} H(W_i,W_j)+\sum_{1\le l<m\le p} \frac{\sqrt{\binom{n}{2}}\mathcal{R}_n^{(l,m)}}{S}=D_{n,1}+D_{n,2},
\end{align*}
where $\mathcal{R}_n^{(l,m)}$ are the remainder terms for $1\le l<m\le p$, and $D_{n,1}$ and $D_{n,2}$ are defined accordingly.
To derive the asymptotic distribution of $D_n$, we use the results from Section \ref{sec:pre} by replacing $U(X_{i,n},X_{j,n})V(Y_{i,n},Y_{j,n})$ with $H(W_i,W_j)$ in Proposition \ref{proposition:hd1}. It provides a neat and convenient way to control the remainder terms in the approximation.
%\begin{remark}
%Another equivalent approach is to notice that each pair of the sample distance covariance $dCov^2_n(W^{(l)},W^{(m)})$ is a U-statistic itself and the Hoeffding decomposition applies; each of the remainder term $\mathcal{R}_n^{(l,m)}$ from decomposition (\ref{eq:u-approx}) is expected to be asymptotically negligible as sample size $n$ goes to infinity. However, the remainder terms for all the pairs need to be controlled simultaneously, which is non-trivial. For simplicity we take the first approach mentioned above.
%\end{remark}
%denote that
%$$ \widetilde D_{n,1}= \frac{1}{S\sqrt{\binom{n}{2}}}\sum_{1\leq i<j\leq n}\{H(W_i,W_j)-E[H(W_i,W_j)]\} $$

First it is straightforward to show that
\begin{align*}
\var(D_{n,1})= &  \frac{1}{S^2 \binom{n}{2}}  \sum_{1\leq i<j\leq n} \sum_{1\leq i'<j'\leq n} \E H(W_i,W_j)H(W_{i'},W_{j'})=\E [H(W,W')^2]/S^2.
%\\ = &  \frac{1}{S^2 \binom{n}{2}}  \sum_{1\leq i<j\leq n} \E H(W_i,W_j)^2
%\\ = & .
\end{align*}
Therefore, we shall choose $S^2=\E [H(W,W')^2]$.
Under the null, $S^2$ can be further simplified as
\begin{align*}
S^2=\sum_{1\le l<m\le p}  dCov^2(W^{(l)})dCov^2(W^{(m)}).
%&\sum_{\substack {1\le l<m\le p \\  1\le l'<m'\le p }} \E \bigg[ U_l(W^{(l)},W'^{(l)})U_m(W^{(m)},W'^{(m)})
%U_{l'}(W^{(l')},W'^{(l')})U_{m'}(W^{(m')},W'^{(m')}) \bigg]
%\\=&\sum_{1\le l<m\le p}  \E U_{l}(W^{(l)},W^{'(l)})^2U_{m}(W^{(m)},W^{'(m)})^2
\end{align*}
Using similar arguments from Section 1.1.1 of the
supplementary material, it can be shown that $D_{n,2}$ is
asymptotically negligible under the null. Based on the above
results, an unbiased estimator for $S^2$ under the null is

\begin{equation}
\label{var}
\hat S^2= \sum_{1\le l<m\le p} dCov^2_n(W^{(l)})dCov^2_n(W^{(m)})
\end{equation}
where $dCov^2_n(W^{(l)})$ is the unbiased estimator for $dCov^2(W^{(l)})$ as defined in Section \ref{sec:pre}.

%\begin{equation}
%\label{var}
%\hat S^2= \sum_{1\le l<m\le p} \left( \frac{1}{\binom{n}{2}}\sum_{1\leq i<j\leq
%n}\widetilde{A}_{ij}^2(l) \right) \left(\frac{1}{\binom{n}{2}}\sum_{1\leq i<j\leq
%n}\widetilde{A}_{ij}^2(m)\right),
%\end{equation}
%where $\widetilde{A}_{ij}(l)$ is the $\mathcal{U}$-centered version of ${A}_{ij}(l)=|W_i^{(l)}-W_j^{(l)}|$.
Therefore, we consider the following feasible test statistic
\[
\hat D_n=\frac{\sum_{1\le l<m\le p}\sqrt{\binom{n}{2}} dCov_n^2(W^{(l)},W^{(m)})}{\hat S}.
\]
In Section~\ref{asy:null}, we establish the asymptotic normality for our test statistic $\hat D_n$ under the null, which leads to the following decision rule for our testing procedure
\[
\phi_{n,\alpha}(W_1,...,W_n) : =\begin{cases}
1 & \mbox{if~}  \hat D_n > z_\alpha\\
0 & \mbox{if~} \hat D_n \le  z_\alpha,\\
\end{cases}
\]
where $z_{\alpha}$ is the $100(1-\alpha)\%$ quantile of standard normal. We reject the null hypothesis $H_0$ if $\phi_{n,\alpha}=1$, do not reject otherwise.

%\begin{remark}
%An alternative variance estimator is given by
%$$
%\hat S^2= \frac{1}{\binom{n}{2}} \sum_{1\leq i<j\leq n} \sum_{1\le
%l<m\le p}\widetilde{A}_{ij}^2(l)\widetilde{A}_{ij}^2(m),
%$$
%where $\widetilde{A}_{ij}(l)$ is defined similar to
%$\widetilde{A}_{ij}$ in Section~\ref{sec:pre} but for the $l$th
%component.
%From (unreported) simulations, we found that these two variance estimators have very comparable performance. %These results also empirically confirm the assumption (\ref{eq:assump3}) imposed later that these two quantities are of the same order.
%Therefore, we only report the results using the one defined in
%(\ref{var}).
%\end{remark}

\subsection{Asymptotic analysis under the null of mutual independence}
\label{asy:null}
To derive the asymptotic distribution for the proposed test statistic $\hat D_n$ under the null, we introduce the following assumptions

\begin{myassumption}{A1}
\label{assumption0}
As $n\rightarrow \infty$ and $p \rightarrow \infty$,
\[
\frac{\sum_{l=1}^{p}\{\E[|W^{(l)}-\mu^{(l)}|]\}^4+n^{-1}\sum_{l=1}^{p}\text{var}(W^{(l)})^2}{ \left\{  \sum_{l=1}^p dCov^2(W^{(l)})\right\}^2}\rightarrow 0,
\]
where $\mu^{(l)}=\E[W^{(l)}]$.

\end{myassumption}
Notice that the first term in assumption \ref{assumption0} can also
be rewritten as $ \sum_{l=1}^p  \{\E [|W^{(l)}-\mu^{(l)}|] \}^4 /
S^2 \rightarrow 0$ as we showed in Section 1.2.2 of the
supplementary material that $\frac{1}{2}[\sum_{l=1}^p
dCov^2(W^{(l)}) ]^2$ is the leading term in the variance, that is,
$S^2 = \frac{1}{2}[\sum_{l=1}^p dCov^2(W^{(l)}) ]^2 \{1 +o(1)\}$
under the null. Therefore, we assume that the sum of the $p$
components' first centered absolute moments to the fourth power
grows at a slower rate than $S^2$, and the sum of the $p$
components' squared variance grows at most $o(n)$ faster than $S^2$.
This is in fact a very mild assumption. For example, when the
element-wise second moments and the distance variances are all lower
and upper bounded uniformly, as in the standard multivariate
Gaussian case, the above assumption is trivially satisfied. Note
that there is no explicit relationship between $p$ and $n$ in the
above assumption, and they are allowed to grow independently.

%Assumption \ref{assumption0} bears a resemblance to  the condition  $\mbox{tr}(\Sigma^4)/\mbox{tr}^2(\Sigma^2)\rightarrow 0 $, as used by \cite{chen2010} in testing high-dimensional sphericity and identity covariance matrices, and in \cite{cq2010} for two sample testing. To connect the two conditions, we can replace the covariance matrix by the distance covariance matrix, where its diagonal elements are the distance variances.

To further appreciate Assumption A1, we mention Assumptions \ref{assumption1}-\ref{assumption2} below, which involve more explicit convergence rate and admit more direct interpretation. It is easy to see that  Assumptions \ref{assumption1}-\ref{assumption2} imply Assumption~\ref{assumption0}, which suffices for our asymptotic analysis.

\begin{myassumption}{B1}
\label{assumption1}
\[
\liminf_{p\rightarrow \infty}\frac{1}{p} \sum_{l=1}^p dCov^2(W^{(l)}) >0. %, \quad \limsup_{p \rightarrow +\infty} \frac{1}{p^2}\sum_{l=1}^p dCov^4(W^{(l)}) = 0
\]
\end{myassumption}

\begin{myassumption}{B2}
\label{assumption2}
\[
\limsup_{p \rightarrow \infty} \frac{1}{p}\sum_{l=1}^p  \var(W^{(l)})^2 < \infty.
\]
\end{myassumption}

Assumption \ref{assumption1} is a mild assumption on the joint distribution of $W$. The inequality sets a lower bound on the average distance variance of the $p$ components of $W$. Notice that $dCov^2(X)=0$ if and only if $X$ is a constant. Therefore, it basically assumes that at least a non-negligible portion of the components of $W$ are not constants.
 %otherwise it is meaningless to conduct the high dimensional independence test on a lot of constants.
Assumption \ref{assumption2} is also fairly mild,  which only requires that the average of squared variance across the $p$ components of $W$ is finite. It is weaker than the assumption that the variance of each component of $W$ is uniformly bounded. %Most of the common distributions can easily satisfy both of the assumptions above.

Proposition 1.1 in the supplementary material provides us a useful tool to derive the asymptotic distribution for our test statistic. We can therefore use the central limit theorem for sum of martingale difference sequences [\cite{hall1984}] to derive the asymptotic distribution for the infeasible test statistic $D_n$, as stated below.

\begin{thm} \label{thm_null}
Under the null hypothesis $H_0$ and Assumption \ref{assumption0}, we have
\[
D_n : = \frac{\sum_{1\le l<m\le p}\sqrt{\binom{n}{2}}
dCov_n^2(W^{(l)},W^{(m)})}{S} \stackrel{d}{\rightarrow} N(0,1),
\]
as $p \rightarrow \infty$ and $n\rightarrow \infty$.
\end{thm}
We obtained the asymptotic normality for the infeasible statistic
$D_n$ without imposing any explicit or implicit constraints on the
growth rates of the dimension $p$ and sample size $n$, and
both can grow to infinity freely. In our feasible test statistic, we
replace $S^2$ by its unbiased estimator $\hat S^2$ as defined in
equation (\ref{var}). We show the ratio consistency of the above
variance estimator in the next theorem.

%\begin{thm}\label{thm_null}
%Define the following quantities,
%\begin{align*}
%\mathcal{V}_1 =&  \E [H(W,W')^2 H(W,W'')^2],\\
%\mathcal{V}_2 =& \E [ H(W,W')H(W,W'')H(W''',W') H(W''',W'')],
%\\\mathcal{V}_3 =&  \E [H(W,W')^4],
%\end{align*}
%where $W',W''$ and $W'''$ are independent copies of $W$. Then under
%the null hypothesis and the assumption that
%\begin{align}
%&\frac{\mathcal{V}_1}{nS^4}\rightarrow 0, \quad \frac{\mathcal{V}_2}{S^4}\rightarrow 0,
%\quad \frac{\mathcal{V}_3}{n^2S^4}\rightarrow 0, \label{eq:con_all}
%\end{align}
%we have
%\begin{align}
%\frac{\sum_{1\le l<m\le p}\sqrt{\binom{n}{2}} dCov_n^2(W^{(l)},W^{(m)})}{S} \rightarrow^d N(0,1).
%\end{align}
%\end{thm}

%\begin{remark}
%Under the null that the $p$ components of $W$ are mutually independent, we can simplify and derive the specific forms for $\mathcal{V}_1$, $\mathcal{V}_2$ and $\mathcal{V}_3$. Please refer to Appendix \ref{sec:3.1} for more details.
%\end{remark}

\begin{thm}
\label{thm_ratio}
Under the null hypothesis $H_0$ and Assumption \ref{assumption0}, we have
\[
\frac{\hat S^2}{S^2} \stackrel{p}{\rightarrow} 1 \quad \mbox{~as~} p \rightarrow \infty \mbox{~and~} n \rightarrow \infty.
\]
\end{thm}
Comparing to Theorem~\ref{thm_null}, we do not impose any additional
assumptions in obtaining the ratio-consistency. Then we can combine
Theorem \ref{thm_null} and Theorem \ref{thm_ratio}, and derive the
asymptotic normality of $\hat D_n$ by applying Slutsky's
theorem.

\begin{corollary}
Under the null hypothesis $H_0$ and Assumption \ref{assumption0}, we have
\[
\hat D_n : = \frac{\sum_{1\le l<m\le p}\sqrt{\binom{n}{2}} dCov_n^2(W^{(l)},W^{(m)})}{\hat S} \stackrel{d}{\rightarrow} N(0,1)
\]
as $p \rightarrow \infty$ and $n\rightarrow \infty$.
\end{corollary}

 It is worth highlighting that our test is developed in a model free setting. No parametric/nonparametric model was assumed and only weak distributional assumptions are required.
  The second moment assumptions seem necessary given the fact that our test is built on sample distance covariance. It is indeed possible to relax the moment assumptions further by
  considering the so-called ranked distance covariance, i.e., replacing sample distance covariance by the sample ranked distance covariance, which is obtained by applying
  distance covariance to the ranks for any two components, say the ranks based on $(W_{1}^{(l)},\cdots,W_{n}^{(l)})$ and $(W_{1}^{(m)},\cdots,W_{n}^{(m)})$, respectively.
  Additionally, it is possible to combine the idea of aggregation with other tests developed for independence of two univariate random variables (see e.g., \cite{heller2013}, \cite{heller2016})
  and form a test for pairwise independence.  These extensions are beyond the scope of this paper and are left
   for future research.

%  free procedure as we do not have any model structure nor distributional assumption on the data, except for the fourth moment constraint, which can be regarded as a weak distributional assumption.

\subsection{Asymptotic analysis under the alternatives}
\label{sec:alternative}
Now we focus on the local alternatives where
some pairs among the $p$ components are dependent, i.e.,
$dCov^2(W^{(l)},W^{(m)}) >0$ for some $l,m \in \{1,...,p\}$. Let
\begin{equation}
D'_n=S^{-1}\sum_{1\le l<m\le p}\sqrt{\binom{n}{2}}
\left\{dCov_n^2(W^{(l)},W^{(m)})-dCov^2(W^{(l)},W^{(m)})\right\},
\end{equation}
where $S^2=\E[H(W,W')^2]$.

By the Hoeffding decomposition, we have
$D_n':=D'_{n,1}+\sum_{1\le l<m\le
p}\sqrt{\binom{n}{2}}\mathcal{R}_n^{(l,m)}/S$, where $D_{n,1}'$ is the leading term
%\begin{align*}
%& =\frac{\sum_{1\le l<m\le p}\sqrt{\binom{n}{2}} \left[dCov_n^2(W^{(l)},W^{(m)})-dCov^2(W^{(l)},W^{(m)})\right]}{S}
%D'_{n,1}:=&\frac{1}{S\sqrt{\binom{n}{2}}}\sum_{1\leq i<j\leq n}\left\{ H(W_i,W_j)-\E H(W_i,W_j)\right\}.
%\\D'_{n,2} :=  &  %\\=& D'_{n,1}+D'_{n,2}
%\end{align*}
%Then we obtain $$
%\sum_{1\le l<m\le p} K_{l,m}(w^{(l)},w^{(m)})=\sum_{1\le l<m\le p}  \E~ U_l(w^{(l)},W^{(l)})~U_m(w^{(m)},W^{(m)})=\E_W(T(W, w, w)) $$
%where $w=(w^{(1)},...,w^{(p)})$ and .
and the contribution from the remainder term
$\sum_{1\le l<m\le p}\sqrt{\binom{n}{2}}\mathcal{R}_n^{(l,m)}/S$ is
asymptotically negligible under the assumptions,
\begin{align}
&\E \left[T(W,W',W'')\right]^2=o(nS^2), \label{eq:assump1}
%\\& \E [T(W,W',W'')T(W',W,W'')]=O(S^2), \label{eq:assump2}
%\\ &\sum_{1\le l<m\le p} dCov^2(W^{(l)})dCov^2(W^{(m)})=O(S^2),
\\ &  \E [\sum_{1\le l<m\le p} U_l(W^{(l)},W'^{(l)}) U_m(W''^{(m)},W'''^{(m)})]^2=o(n^2S^2) ,  \label{eq:assump3}
\\&\var\left( \E_W(T(W, W', W'))  \right)=o(n^{-1}S^2),\quad
\var\left(\E_W(T(W, W', W''))\right)=o(S^2),
\label{eq:assump4}
\end{align}
where we define %$K_{l,m}(W^{(l)},W^{(m)})=\E[U_l(W^{(l)},W'^{(l)})U_m(W^{(m)},W'^{(m)})|W]$ and
$T(W,W',W'')=\sum_{1\le l<m\le p} U_l(W^{(l)},W^{''(l)}) U_m(W^{(m)},W^{'(m)})$ and $\E_W$ denotes the expectation with respect to $W$.

Conditions (\ref{eq:assump1})--(\ref{eq:assump4}) are
obtained from (\ref{eq:assumption1})--(\ref{eq-alt}) and they characterize the local alternative we discuss here in an abstract way. Notice that under the null of mutual independence, these conditions are automatically satisfied and $\var\left( \E_W(T(W, W', W'))  \right) = 0$ in Condition (\ref{eq:assump4}), which makes our test statistic a degenerate U-statistic under the null. For the local alternative, we also focus on the degenerate case in the sense that we require the alternative not too far away from the null. Therefore, these conditions guarantee that our test statistic is still degenerate when some pairs among the $p$ components are dependent.  In the case that $\var\left( \E_W(T(W, W', W'))  \right) $ does not vanish and the test statistic is non-degenerate, we can regard it as the fixed alternative; its asymptotic distribution can be derived similarly under suitable assumptions.
%{(\color{red} ? In particular, Condition (\ref{eq:assump4}) has a similar interpretation as given in Remark 2.4 of our MDD paper.)

Furthermore, we can rewrite $D'_{n,1}$ under Condition (\ref{eq:assump4}) using the double centered version of $H(W,W')$ as
\begin{align*}
D'_{n,1} =& \frac{1}{S\sqrt{\binom{n}{2}}}\sum_{1\leq i<j\leq n}
\widetilde H(W_i,W_j)+o_p(1),
%\\=& \frac{1}{S\sqrt{\binom{n}{2}}}\sum_{1\leq i<j\leq n}  \widetilde H(W_i,W_j)  +o_p(1)
\end{align*}
where $\widetilde H(W,W')=H(W,W')-\E [H(W,W')|W]-\E [H(W,W')|W']+ \E [H(W,W')]$. Similar to the arguments under the null in Section \ref{asy:null} and Propositions 1.2-1.3 in the supplementary material, we define the following quantities
\begin{align*}
& \mathcal{\widetilde V}_1 =  \E [\widetilde H(W,W')^2 \widetilde
H(W,W'')^2],\quad \mathcal{\widetilde V}_2 = \E [ \widetilde
H(W,W')\widetilde H(W,W'')\widetilde H(W''',W') \widetilde
H(W''',W'')],
\\ & \mathcal{\widetilde V}_3 =  \E [\widetilde H(W,W')^4],
\quad \widetilde S^2= \E[ \widetilde H(W,W')^2].
\end{align*}

The following theorem establishes the asymptotic normality for $D'_n$ based on similar arguments under the null with $H$ replaced by $\widetilde H$.
\begin{thm}\label{thm_alt}
Under the Conditions (\ref{eq:assump1})--(\ref{eq:assump4}) and also
\begin{align}
\label{eq:clt_condition}
&\frac{\mathcal{\widetilde V}_1}{n\widetilde S^4}\rightarrow 0, \quad \frac{\mathcal{\widetilde V}_2}{\widetilde S^4}\rightarrow 0,
\quad \frac{\mathcal{\widetilde V}_3}{n^2\widetilde S^4}\rightarrow 0,
\end{align}
we have $D'_n\rightarrow^d N(0,1).$
\end{thm}

Using Theorem \ref{thm_alt}, we can readily show that the power function of the test statistic $\hat D_n$ is approximately
\begin{align*}
& \Phi\left (-z_{\alpha}+ \sqrt{\binom{n}{2}} \sum_{1\le l<m\le p} dCov^2(W^{(l)},W^{(m)})/S \right )
%\\=&\Phi\left (-z_{\alpha}+ \sqrt{\binom{n}{2}} \frac{\sum_{1\le l<m\le p} dCov^2(W^{(l)},W^{(m)})}{\sqrt{\sum_{1\le l<m\le p}  \E U_{l}(W^{(l)},W^{'(l)})^2U_{m}(W^{(m)},W^{'(m)})^2}} \right ),
\end{align*}
where $\Phi(\cdot)$ and $z_{\alpha}$ are the distribution function and $100(1-\alpha)\%$ quantile of standard normal respectively. % {\color{red} If $\sum_{1\le l<m\le p} dCov^2(W^{(l)},W^{(m)}) =O(n^{-1}S)$, for example, we have a non-trivial power function approximately $\Phi\left (-z_{\alpha}+ O(1)\right)$.}
%when $\hat S$ is an appropriate estimator of $S$,

\subsection{Rate optimality under Gaussian equicorrelation}
When the dependence is  weak, it may be  difficult to distinguish
between the null and the alternative hypothesis. In this subsection,
we study the boundary for the testable, non-testable region and
conduct power analysis for our test from a minimax point of view
following the work of \cite{Cai2013}, and also show that our test is
rate optimal. We focus on the case where $W=(W^{(1)},...,W^{(p)})$
follows a $p$-variate Gaussian distribution. Without loss of
generality, we assume each of the marginals is standard Gaussian
with unit variance. Then our null hypothesis is equivalent to
$\Sigma -I_p = 0$.  We introduce the following alternative class
${\cal N}_p( ||\Sigma - I_p||_F \ge c)$ which was also discussed in
\cite{Cai2013}, \cite{leung2015},
\[
{\cal N}_p( ||\Sigma - I_p||_F \ge c) : = \{W=(W^{(1)},...,W^{(p)})| W \sim N_p(\mu, \Sigma),~ ||\Sigma - I_p||_F \ge c  \}
\]
where $N_p(\mu, \Sigma)$ denotes a $p$-variate Gaussian distribution
with mean $\mu$ and covariance matrix $\Sigma$, $||\cdot || _F$ is
the matrix Frobenius norm and $I_p$ is the $p
 \times p$ identity matrix. Here $||\Sigma - I_p||_F$ quantifies the signal/dependence strength and the difficulty of the testing problem depends on $c$. A similar alternative class is also discussed in \cite{han2014} based on the maximum norm.

 Theorem 1 of \cite{Cai2013} shows that under the regime that $p/n$ is bounded, for sufficiently small $b$ such that $c=b \sqrt{p/n}$,
 no level-$\alpha$ test can distinguish between the null and alternative with desired power, that is, for any generic level-$\alpha$ test $\varphi$  and $0<\alpha<\beta<1$,
 $$\limsup_{n \rightarrow \infty} \inf_{{\cal N}_p( ||\Sigma - I_p||_F \ge c)  } \E(\varphi) < \beta.$$
 Therefore, $c=b \sqrt{p/n}$ sets the lower bound for the separation rate between the null and alternative in order for any test to distinguish between them. If a test can achieve arbitrary large power for large enough $ b^*$, i.e.,  for any $0<\alpha<\beta^*<1$, we can have
  $$\liminf_{n \rightarrow \infty} \inf_{{\cal N}_p( ||\Sigma - I_p||_F \ge c^*)  } \E(\varphi) > \beta^*$$
for  $c^* =  b^* \sqrt{p/n}$, then the test is called rate optimal. More discussion can be found in \cite{Cai2013} and \cite{leung2015}.
Below we show the rate optimality of our proposed test.

 Consider the equicorrelation alternative class ${\cal N}_p^{equi}( ||\Sigma - I_p||_F \ge c) $, which is a sub-class of ${\cal N}_p$ such that all the pairwise correlations equal to a common value denoted as $\rho$. Let $\Theta = (dCov(W^{(l)},W^{(m)}))_{1\le l < m \le p}$ be the  $p \choose 2$-vector of all the pairwise distance covariance.
 It is easy to see that ${\cal N}_p^{equi}( ||\Sigma - I_p||_F \ge c) $ is equivalent to ${\cal N}_p^{equi}( |\Theta | \ge \tilde c) $ for some $\tilde{c}$. Here we use the fact that for standard Gaussian variables with correlation $\rho$, we have
\begin{equation}
dCov^2(W^{(l)},W^{(m)}) = \frac{4}{\pi} [ \rho \arcsin \rho  + \sqrt{1-\rho^2} -\rho \arcsin (\rho/2) - \sqrt{4-\rho^2} +1 ] := f(\rho).
\end{equation}
In view of the proof of Theorem 7 in \cite{szekely2007}, we have
that $ c_1 \rho^2 p^2 \le  |\Theta |^2 =  \sum_{l<m}
dCov^2(W^{(l)},W^{(m)}) \le   c_2 \rho^2p^2 $ for some positive
constants $c_1$ and $c_2$. With a slight abuse of notation, we shall
use $\phi_{n,\alpha}$ to denote the decision rule based on our
infeasible test statistic $D_n$ in this subsection.

\begin{thm} \label{thm:rate}
For any $0 < \alpha < \beta < 1$, as $p/n \rightarrow \lambda \in (0, \infty) $, there exists a constant $\tilde c = \tilde c(\alpha, \beta, \lambda) > 0 $ such that
\[\liminf_{n \rightarrow \infty} \inf_{{\cal N}_p^{equi}( |\Theta | \ge \tilde c)  } \E[\phi_{n,\alpha}] > \beta. \]
\end{thm}

We conjecture that the same result presented in Theorem~\ref{thm:rate} also holds for the feasible test statistic $\hat{D}_n$, but it seems very involved
to derive a probabilistic bound for $\hat{S}^2/S^2-1$, which is required in the proof. Nevertheless, the above result suggests that our distance covariance based
test has potentially good power properties in the special case of Gaussian distributions, as shared by rank correlation based test of \cite{leung2015}.
See Section~\ref{sec:sim} for numerical evidence.

\section{Testing for Banded Dependence Structure}
\label{sec:extend}
We propose a test statistic in this section to test for the banded dependence ($m$-dependence) structure. Usually when the null hypothesis of mutual independence is rejected, it is of interest to test for some specific dependence structure afterwards or independently. For example, when the $p$ components have a natural ordering, which arises in time series analysis, testing for $m$-dependence is of particular interest [see \cite{moon2013}]. Moreover, in the high dimensional covariance matrix estimation literature, banded covariance structure attracts a lot of attention; see \cite{wu2003}, \cite{bickel2008}, \cite{wagaman2009}, \cite{shao2014} among others. \cite{qiu2012} built a test for banded covariance matrices and also presented an approach to estimating the corresponding bandwidth; \cite{cai2011}, \cite{han2014} used Pearson correlation and rank correlation respectively for testing banded linear and monotone dependence. In contrast, our proposed test for banded structure (or $m$-dependence structure) targets any kind of dependence using distance covariance as analogous to the mutual independence test in Section \ref{sec:main}. Accordingly, we consider the following null hypothesis for the banded dependence structure:

\[H_{0,h}: W^{(l)} \text{~and~} W^{(m)}~\mbox{are independent for all~} |l-m|\ge h.\]
Define
\[ H_h(W_i,W_j)= \sum_{\substack{1\le l<m\le p\\ |l-m|\ge h }}U_{l}(W_i^{(l)},W_j^{(l)})U_{m}(W_i^{(m)},W_j^{(m)}).\]
Then the (infeasible) distance covariance based statistic for testing $H_{0,h}$ is
\[
D_{n,h}=S_h^{-1}\sum_{\substack{1\le l<m\le p\\ |l-m|\ge h }}\sqrt{\binom{n}{2}} dCov_n^2(W^{(l)},W^{(m)}),
\]
where $S_h^2=\E [H_h(W,W')^2]$. The variance estimator we consider
here is
\begin{equation*}
\hat S_h^2= \frac{1}{\binom{n}{2}}\sum_{1\leq i<j\leq
n}  \left( \sum_{\substack{1\le l<m\le p\\ |l-m|\ge h }}  \widetilde{A}_{ij}(l) \widetilde{A}_{ij}(m) \right)^2,
\end{equation*}
where $\widetilde{A}_{ij}(l)$ is defined similar to $\widetilde{A}_{ij}$ but is based on the data $\{W_i^{(l)}\}_{i=1}^{n}$.
Similarly define $\widetilde S_h^2$, $\mathcal{V}_{j,h}$ and $\mathcal{\widetilde V}_{j,h}$ as the $h$-lag analogues of $\widetilde S^2$, $\mathcal{V}_{j}$ and
$\mathcal{\widetilde V}_{j}$ for $1\leq j\leq 3$ from Section
\ref{sec:main}. Following similar arguments in the proofs of Theorem
\ref{thm_null} and Theorem \ref{thm_alt}, we have the following theorem for testing banded dependence structure.

%\begin{thm}\label{thm:depend_null}
%Under the assumption that
%\begin{align*} &\frac{\mathcal{V}_{1,h}}{nS_h^4}\rightarrow 0, \quad \frac{\mathcal{V}_{2,h}}{S_h^4}\rightarrow 0, \quad \frac{\mathcal{V}_{3,h}}{n^2S_h^4}\rightarrow 0, \label{eq:con_all_m}
%\end{align*}
%and the null hypothesis, we have $D_{n,h}\rightarrow^d N(0,1).$
%\end{thm}

\begin{thm}\label{thm:depend}
Define $T_h(W,W',W'')=\sum_{\substack{1\le l<m\le p\\|l-m|\ge h}} U_l(W^{(l)},W^{''(l)}) U_m(W^{(m)},W^{'(m)})$. Then under the assumptions that
\begin{align}
&\E \left[T_h(W,W',W'')\right]^2=o({nS_h^2}), \label{eq:bandassumption1}
%\\& \E [T_h(W,W',W'')T_h(W',W,W'')]=O(S_h^2),
\\& \E [  \sum_{\substack{1\le l<m\le p\\|l-m|\ge h}} U_l(W^{(l)},W^{'(l)}) U_m(W''^{(m)},W'''^{(m)})
]^2 =o({n^2S_h^2}), \label{eq:bandassumption2}
%\\& \sum_{\substack{1\le l<m\le p\\|l-m|\ge h}}dCov^2(W^{(l)}) dCov^2(W^{(m)})
\\ & \var\left( \E_W(T_h(W, W', W')) \right)=o(n^{-1}S_h^2), \quad \var\left(\E_W(T_h(W, W',
W'')\right)=o(S_h^2) \label{band-t}
\end{align}
and also
\begin{align}
&\frac{\mathcal{\widetilde V}_{1,h}}{n\widetilde S_h^4}\rightarrow 0, \quad \frac{\mathcal{\widetilde V}_{2,h}}{\widetilde S_h^4}\rightarrow 0,
\quad \frac{\mathcal{\widetilde V}_{3,h}}{n^2 \widetilde S_h^4}\rightarrow 0,
\end{align}
we have
\begin{align*}
D_{n,h}' : = S_h^{-1}\sum_{\substack{1\le l<m\le p\\|l-m|\ge
h}}\sqrt{\binom{n}{2}}
\left[dCov_n^2(W^{(l)},W^{(m)})-dCov^2(W^{(l)},W^{(m)})\right]
\rightarrow^d N(0,1).
\end{align*}
Furthermore, under the null hypothesis of banded dependence,
$\E_W(T_h(W, W', W'))=0$ and condition (\ref{band-t}) is satisfied
automatically; $\widetilde S_h^2$, $\mathcal{\widetilde V}_{j,h}$
reduce to $ S_h^2$ and $\mathcal{V}_{j,h}$ for $1 \le j \le 3$. We
have $D_{n,h}\rightarrow^d N(0,1).$
\end{thm}

Similar to the discussion in Section \ref{sec:alternative}, the
theorem is presented under an abstract local alternative class
characterized by (\ref{eq:bandassumption1}) - (\ref{band-t}). These
conditions can be further studied under a more specific definition
of the local alternative class, which we did not pursue here in this
paper.

\section{Joint Dependence Metrics}
\label{sec:joint}

Although most test statistics aimed for mutual independence in the literature only target at the pairwise independence, there are more ambitious tests that quantify the overall joint dependence directly, see e.g. \cite{kankainen1995} and \cite{Pfister:2016aa}. The latter proposed the dHSIC as an extension of the two variable Hilbert-Schmidt independence criterion  (HSIC) [\cite{gretton2005}, \cite{gretton2007}, \cite{smola2007}] to the multivariate case. It embeds the joint distribution and the product of the marginal distributions into a reproducing kernel Hilbert space (RKHS, hereafter) and measures their squared distance. Following the notations from \cite{Pfister:2016aa}, we denote ${\mathbb P}^{(W^{(1)},...,W^{(p)})}$  the joint probability distribution for $W$, which is a $p$-dimensional random vector $W= (W^{(1)},...,W^{(p)} )$, and ${\mathbb P}^{W^{(i)}}$ the marginal probability distribution for $W^{(i)}$. Let $(W_j)_{j \in \mathbb N}$ be a sequence of iid copies of $W$.  Let $k_i$ be a continuous, bounded , positive semi-definite kernel associated with $W^{(i)}$ and denote by ${\cal H}_i$ the corresponding RKHS. Further denote $k = k_1 \otimes ... \otimes k_p $ the tensor product of the kernels $k_i$ and ${\cal H} = {\cal H}_1 \otimes ...\otimes {\cal H}_p$ the tensor product of the RKHSs ${\cal H}_i$. Let $\Pi(\cdot)$ be the mean embedding function associated with $k$. Then the dHSIC is defined as
\begin{align*}
\mbox{dHSIC}(W) : = & \Big|\Big| \Pi\Big({\mathbb P}^{W^{(1)}} \otimes ... \otimes {\mathbb P}^{W^{(p)}} \Big) - \Pi\Big({\mathbb P}^{(W^{(1)},...,W^{(p)})}\Big) \Big|\Big|^2_{\cal H}\\
= & \E \left(  \prod_{i=1}^p k_i(W^{(i)}_1, W^{(i)}_2 )   \right) +
\E \left(  \prod_{i=1}^p k_i(W^{(i)}_{2i-1}, W^{(i)}_{2i} )
\right)  - 2 \E \left(  \prod_{i=1}^p k_i(W^{(i)}_1, W^{(i)}_{i+1} )
\right).
\end{align*}

 As long as the kernel is characteristic, the embedding of Borel probability measures will be injective, which implies that the squared distance above is zero if and only if the joint distribution is the same as the product of the marginal distributions. A commonly used kernel is the Gaussian kernel $k(w,w') = \exp\{-||w-w'||^2/(2\gamma^2)\}$, which is characteristic but contains a bandwith parameter $\gamma$.

Another joint dependence measure was proposed by \cite{kankainen1995}. This metric is also based on characteristic functions. It is a weighted integral over the difference between the joint characteristic function and the product of the marginal characteristic functions. In \cite{Pfister:2016aa}, they showed that this dependence metric is a special case of the dHSIC by choosing a specific kernel function as defined in equation (2.5) of the latter paper. However, sample version for dHSIC  is only defined when the sample size is at least twice as large as the dimension of the data, which makes it unsuitable for the setting where the dimension  exceeds the sample size. It is also unclear how the dHSIC-based test performs when the dimension is relatively large compared to sample size, as the dimension in all simulation examples in \cite{Pfister:2016aa} are small and their theory is only for the fixed-dimensional case.
We shall examine the finite sample performance in the high dimensional setting in Section~\ref{sec:sim_mutual}.

%From the simulation results in Table 6, our conjecture is that as the dimension increases the joint dependence metric will suffer from decreasing power.

On the other hand, due to the equivalence of distance-based and
RKHS-based statistics as discussed in \cite{sejdinovic2013}, we can
borrow the idea of embedding mentioned above and extend the distance
covariance to construct a multivariate distance covariance as an
alternative. Again consider the random variables
$W=(W^{(1)},W^{(2)},\dots,W^{(p)})$, where $W^{(i)} \in {\mathbb
R}^{d_i}$. Here we allow each $W^{(i)}$ to have different
dimensions. For $0<a<2$, there exists an embedding
$w\rightarrow\phi(,w):\mathbb{R}^d\rightarrow \mathcal{H}_d$ such
that for any $w,w'\in\mathbb{R}^d$,
\begin{align}\label{eq1}
<\phi(\cdot,w),\phi(\cdot,w')>_{\mathcal{H}_d}=|w|^{a}+|w'|^{a}-|w-w'|^{a} : =\mathcal{K}(w,w';a),
\end{align}
where $\mathcal{K}$ is the so-called distance induced kernel
associated with the Euclidean norm, and $\mathcal{H}_d$ is some
Hilbert space with the inner product
$<\cdot,\cdot>_{\mathcal{H}_d}$ [see Proposition 3 of \cite{sejdinovic2013}]. For instance, when $a=1$, one
can choose $\mathcal{H}_d=L^2(\mathbb{R}^d,\lambda^d)$ with
$\lambda^d$ being the Lebesgue measure and
$\phi(w,w')=\{|w-w'|^{-(d-1)/2}-|w'|^{-(d-1)/2}\}/c_0$ for some
constant $c_0>0$, see \cite{lyons2013}. Denote by $\mathcal{K}_i$ the
distance induced kernel associated with $\mathbb{R}^{d_i}$.
The multivariate distance covariance (MdCov) is defined as
{\footnotesize \begin{align*}
&MdCov^2(W;a) =\E  \left( \prod_{i=1}^{p}\mathcal{K}_i(W_1^{(i)},W_2^{(i)};a) \right) + \E \left( \prod^{p}_{i=1} \mathcal{K}_i(W_{2i-1}^{(i)},W_{2i}^{(i)};a) \right)-2\E \left(\prod_{i=1}^{p}\mathcal{K}_i(W_1^{(i)},X_{i+1}^{(i)};a)\right).
\end{align*}}
%Thus we have $||\phi(,x)-\phi(,x')||^2=|x-x'|$, where $||\cdot||^2$ denotes the inner product induced norm.
To understand MdCov, we consider the tensor embedding
$(w^{(1)},\dots,w^{(p)})\rightarrow \phi_1(,w^{(1)})\times \cdots \times
\phi_d(,w^{(p)}):\mathbb{R}^{d_1}\times \cdots \times
\mathbb{R}^{d_p}\rightarrow
\widetilde{\mathcal{H}}:=\mathcal{H}_{d_1}\times
\cdots\times \mathcal{H}_{d_p}$, where $\phi_i$ is the embedding associated
with $\mathbb{R}^{d_i}.$ The inner product on
$\widetilde{\mathcal{H}}$ satisfies that $<h_1\otimes\cdots \otimes
h_p,g_1\otimes\cdots \otimes
g_p>_{\widetilde{\mathcal{H}}}=\prod_{i=1}^{p}<h_i,g_i>_{\mathcal{H}_{d_i}}$.
%Denote by $\boldsymbol \mu $ the joint measure for $(W^{(1)}, W^{(2)},\dots,W^{(p)})$ and
%$\mu_i$ the marginal measure associated with $W^{(i)}$.
 Let $\mathfrak{M}$ denote the space of measures on
$\mathbb{R}^{d_1}\times \cdots \times \mathbb{R}^{d_p}$. Define the map on
$\mathfrak{M}$,
$$\Pi^{\cal K} (\boldsymbol {\mathbb P}^{(W^{(1)},...,W^{(p)})} )=\int \prod_{i=1}^{p}\phi_i(\cdot,w^{(i)})d {\mathbb P}^{(W^{(1)},...,W^{(p)})} .$$
%$$\beta_{\phi_1\otimes \cdots\otimes \phi_p}(\boldsymbol \mu )=\int \prod_{i=1}^{p}\phi_i(\cdot,w^{(i)})d\boldsymbol \mu ((w^{(1)},\dots,w^{(p)})).$$
The following result provides an equivalent representation for
MdCov.
\begin{proposition}\label{prop8}
For $0 < a < 2$, assume that $ E|W^{(i_1)}|^{a}|W^{(i_2)}|^{a}\cdots |W^{i(_m)}|^{a}<\infty$ for any $m$-tuple $(i_1,\dots,i_m)$ and $1\leq m\leq p$. Then
$$Mdcov^2(W;a)=\Big|\Big|\Pi^{\cal K} \left(\boldsymbol {\mathbb P}^{(W^{(1)},...,W^{(p)})} \right)  - \Pi^{\cal K} \left( {\mathbb P}^{W^{(1)}} \otimes ... \otimes {\mathbb P}^{W^{(p)}} \right)  \Big|\Big|_{\widetilde{\mathcal{H}}}^2.$$
%$$Mdcov^2(W^{(1)},W^{(2)},\dots,W^{(p)};\alpha)=||\beta_{\phi_1\otimes \cdots\otimes \phi_p}({\mathbb P}^{(W^{(1)},...,W^{(p)})} ) -{\mathbb P}^{W^{(1)}} \otimes ... \otimes {\mathbb P}^{W^{(p)}} )||^2.$$
\end{proposition}

MdCov can be viewed as another special case of the dHSIC in
\cite{Pfister:2016aa} and \cite{sejdinovic2013nips}, where the
kernel is chosen to be the distance induced kernel $\mathcal{K}$.
And for this particular kernel, there is no bandwidth parameter that
needs to be tuned but one has to select the parameter $a$.
Below we also point out a characteristic function interpretation for
MdCov based on the Fourier embedding. Let
$c_{a,d}=2\pi^{d/2}\Gamma(1-a/2)/\{a2^{a}\Gamma((d+a)/2)\}.$
Denote $d\tilde t=(c_{a,d_1}c_{a,d_2}\dots
c_{a,d_p}|t_1|_{d_1}^{a+d_1}\cdots|t_p|_{d_p}^{a+d_p}
)^{-1}dt_1\cdots dt_p$ and $f_i(t_i) = \E e^{\I \langle
t_i,W^{(i)}\rangle}$ as the characteristic function for $W^{(i)}$.

\begin{proposition}\label{prop77}
MdCov can be rewritten as
\begin{align*}\label{eq-fourier}
MdCov^2(W;a)=&\int\left|\E\left[\prod^{d}_{i=1}(e^{\I
\langle t_i,W^{(i)}\rangle}-1)\right]-\prod^{d}_{i=1}(f_i(t_i)-1)\right|^2d\tilde t.
\end{align*}
\end{proposition}

Notice that the above definition is slightly different from the dependence measure proposed in \cite{kankainen1995} since they correspond to different kernels. Surprisingly, different from the dHSIC in \cite{Pfister:2016aa}, the MdCov does not completely
characterize mutual independence. To see this, suppose $X_1$ has a
degenerate distribution at zero, and $(X_2,\dots,X_d)$ are dependent.
Then the MdCov is equal to zero while
$(X_1,\dots,X_d)$ are dependent. This is essentially due to the fact
that the barycenter map associated with the tensor product space for
dimension greater than two [\cite{lyons2013}] is no longer injective.

Similar to \cite{Pfister:2016aa},  for a set of random samples $W_1,...,W_n$, the sample version for MdCov can be defined by a $V$-statistic,
{\small
\begin{align*}
 MdCov_n^2(W;a) =&  \frac{1}{n^2} \sum_{M_{2,n}}  \left( \prod_{i=1}^{p}\mathcal{K}_i(W_{k_1}^{(i)},W_{k_2}^{(i)};a) \right)   +  \frac{1}{n^{2p}} \sum_{M_{2p,n}} \left( \prod^{p}_{i=1} \mathcal{K}_i(W_{k_{2i-1}}^{(i)},W_{k_{2i}}^{(i)};a) \right)\\
& -\frac{2}{n^{p+1}} \sum_{M_{p+1,n}} \left(\prod_{i=1}^{p}\mathcal{K}_i(W_{k_1}^{(i)},X_{k_{i+1}}^{(i)};a)\right).
\end{align*}
}
where $M_{q,n} = \{1,...,n\}^q$ is the $q$-fold Cartesian product of the set $\{1,...,n\}$ and $(k_1,...,k_q) \in M_{q,n} $ for $n\in\{2p,2p+1,...\}$. The implementation of MdCov-based test is similar to that of dHSIC,  in that we can easily conduct a permutation test based on the random sample.

\section{Simulation}
\label{sec:sim}
In this section, we conduct Monte Carlo simulations to assess the finite sample performance of the mutual independence test in Section \ref{sec:sim_mutual}, and present a comparison between pairwise independence test with a joint independence test in Section \ref{sec:buhlmann}; we also compare our proposed methods (dCov, hereafter) with the following existing tests in the literature. \cite{schott2005} proposed a ${\mathcal L}_2$ type statistic using pairwise Pearson correlation (SC, hereafter); \cite{leung2015} studied the ${\mathcal L}_2$ type statistics using Kendall's tau ($\text{LD}_\tau$), Spearman's rho ($\text{LD}_\rho$) or the sign covariance introduced by \cite{bergsma2014} ($\text{LD}_{t^*}$); Along a different line \cite{cai2011} used the ${\mathcal L}_\infty$ type statistic of Pearson Correlation to test the structure of covariance matrices (CJ test, hereafter); \cite{han2014} developed the ${\mathcal L}_\infty$ type statistics using either Kendall's tau ($\text{HL}_\tau$) or Spearman's rho ($\text{HL}_\rho$). Section \ref{sec:buhlmann} further compared the joint independence tests dHSIC and MdCov discussed in Section \ref{sec:joint} with our proposed test.
Some additional simulation results for testing mutual independence and banded dependence structure are presented in Section 2 of the supplementary material.

\subsection{Testing for mutual independence}
\label{sec:sim_mutual}

In this subsection, we evaluate the size and power of the proposed mutual independence test for both  Gaussian and non-Gaussian distributions. The size and power (rejection probabilities) reported below are based on 5000 Monte Carlo simulations at the nominal level $\alpha=0.05$. We choose sample size $n=\{60,100\}$ and the dimension $p=\{50,100,200,400,800\}$.

\begin{example}
\label{eg:size}
 The data $W=(W_1,...,W_p) \in {\mathbb R}^p$ are generated as follows with each component independent from others
\begin{itemize}
\item {\bf i)} The data are generated from a standard Gaussian distribution with $W \sim N_p(0, I_p)$;
\item {\bf ii)} The data are generated from a Gaussian copula family with $W=Z^{1/3}$ and $Z \sim N_p(0, I_p)$;
\item {\bf iii)} The data are generated from a Gaussian copula family with $W=Z^3$ and $Z \sim N_p(0, I_p)$;
\item {\bf iv)} The components $\{W_j\}_{j=1}^p$ are i.i.d. from the student-$t$ distribution with degrees of freedom three.
\end{itemize}
\end{example}

\begin{figure}[h!]
\centering
\includegraphics[width=4.3in]{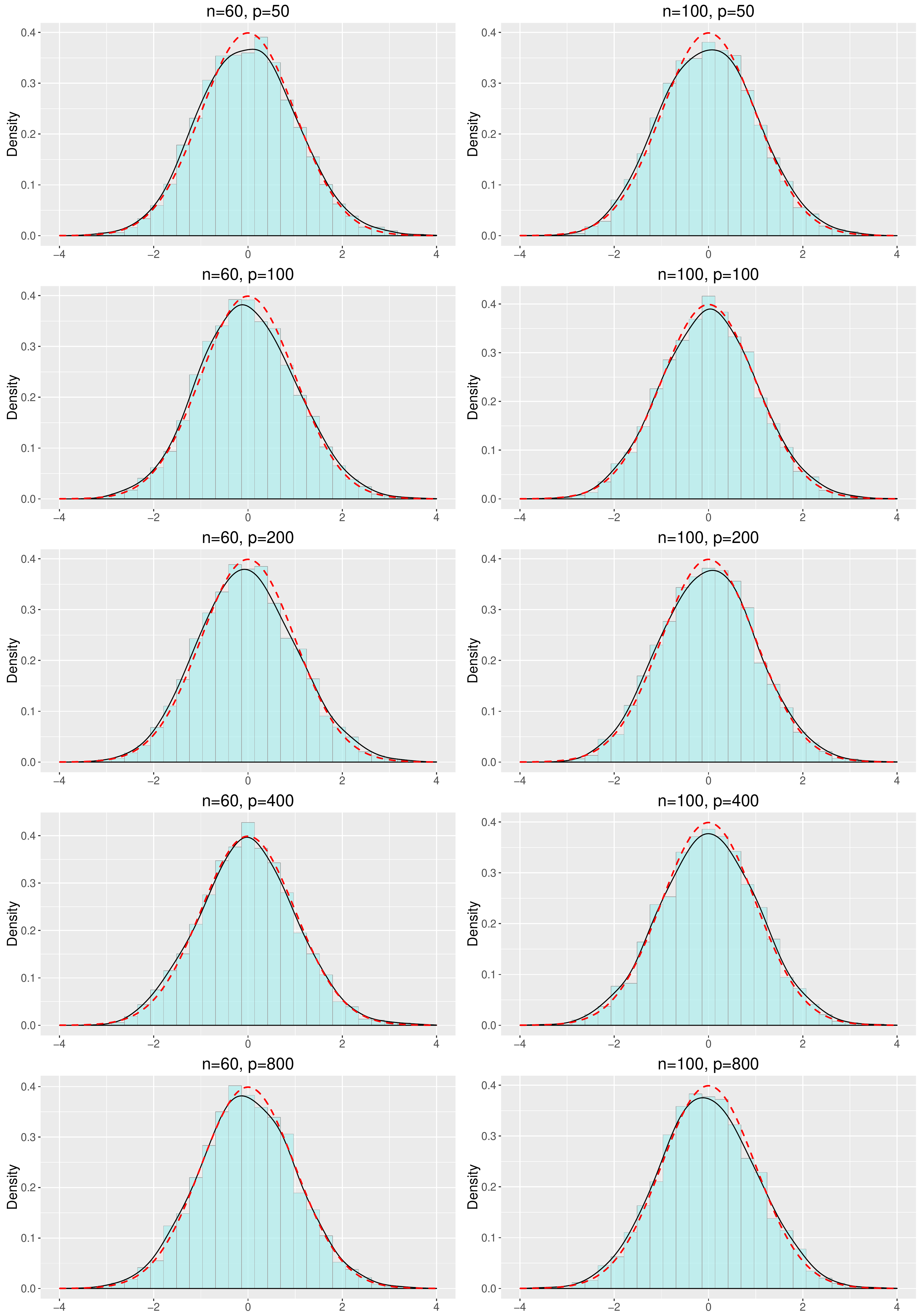}
\caption{\small The histogram and kernel density estimate for the null distribution of the test statistics for Example \ref{eg:size}. The red dashed line is the density of the standard normal.}
\label{fig:size}
\end{figure}

The sizes for all the tests are summarized in Table \ref{table:size}. The performance of the proposed test is very comparable to those from $\text{LD}_\tau$ and $\text{LD}_\rho$. % although it has slightly inflated size when $n=60$.
SC's test performs reasonably well in cases i) and ii), especially when the underlying data is Gaussian. However, it has slightly upward size inflation in case iv) and exhibits severe size distortion in case iii). We also observed some size inflation from $\text{LD}_{t^*}$ when the sample size is small. The ${\mathcal L}_\infty$ type statistics $\text{HL}_\tau$ and $\text{HL}_\rho$ turn out to be conservative for all the scenarios; CJ's test has an unpleasantly high rejection rate in cases iii) and iv) due to the violation of Gaussian assumption. In addition, Figure \ref{fig:size} shows the histogram of the dCov-based test statistic from 5000 Monte Carlo simulation of case i) as well as the kernel density estimate using the Gaussian kernel. Comparing with the red dashed line (density of standard normal), we observe that the null distribution of our test statistic is in general very close to standard normal for all the combinations of $(n, p)$ being considered.

\begin{table}[h]
\scriptsize
\centering
\caption{Size of the tests from Example \ref{eg:size} }
\resizebox{\columnwidth}{!}{%
\label{table:size}
\begin{tabular}{cccccccccccccccccc}
  \hline
  && \multicolumn{8}{c}{\bf (i)} & \multicolumn{8}{c}{\bf (ii)} \\
\cmidrule(lr){3-10} \cmidrule(lr){11-18}
  $n$ & $p$ & dCov & SC & $\text{LD}_\tau$ & $\text{LD}_\rho$ & $\text{LD}_{t^*}$ & CJ & $\text{HL}_\tau$ & $\text{HL}_\rho$  & dCov & SC & $\text{LD}_\tau$ & $\text{LD}_\rho$& $\text{LD}_{t^*}$  & CJ & $\text{HL}_\tau$ & $\text{HL}_\rho$ \\
  \hline
   60 &   50 & 0.054 & 0.047 & 0.061 & 0.050 & 0.066&  0.016 & 0.030 & 0.017 & 0.057 & 0.053 & 0.062 & 0.054 & 0.077 &  0.028 & 0.027 & 0.016 \\
     60 &  100 & 0.054 & 0.049 & 0.062 & 0.051 & 0.074 & 0.010 & 0.026 & 0.010 & 0.052 & 0.050 & 0.060 & 0.050 & 0.075 &  0.024 & 0.025 & 0.010 \\
      60 &  200 & 0.056 & 0.056 & 0.065 & 0.056 & 0.068& 0.004 & 0.021 & 0.007 & 0.057 & 0.055 & 0.060 & 0.049 & 0.067 & 0.017 & 0.021 & 0.006 \\
     60 &  400 & 0.050 & 0.046 & 0.060 & 0.048 &  0.067& 0.001 & 0.018 & 0.002 & 0.058 & 0.054 & 0.062 & 0.050  & 0.074 & 0.014 & 0.019 & 0.005 \\
      60 &  800 & 0.048 & 0.040 & 0.052 & 0.045 &  0.060 & 0.000 & 0.011 & 0.001 & 0.061 & 0.055 & 0.057 & 0.050 & 0.064 & 0.009 & 0.014 & 0.002 \\
     100 &   50 & 0.054 & 0.049 & 0.054 & 0.052 & 0.059  & 0.021 & 0.033 & 0.021 & 0.051 & 0.050 & 0.056 & 0.049 & 0.069 & 0.033 & 0.031 & 0.024 \\
    100 &  100 & 0.054 & 0.054 & 0.057 & 0.050 & 0.066 & 0.018 & 0.032 & 0.021 & 0.048 & 0.046 & 0.052 & 0.046 & 0.053 & 0.031 & 0.033 & 0.022 \\
    100 &  200 & 0.053 & 0.050 & 0.054 & 0.049 & 0.066 & 0.013 & 0.030 & 0.018 & 0.046 & 0.045 & 0.050 & 0.044 & 0.064 & 0.032 & 0.032 & 0.020 \\
     100 &  400 & 0.055 & 0.047 & 0.054 & 0.051&  0.061 & 0.012 & 0.027 & 0.013 & 0.048 & 0.049 & 0.050 & 0.046 & 0.054 & 0.022 & 0.026 & 0.013 \\
     100 &  800 & 0.057 & 0.052 & 0.060 & 0.056 & 0.062 & 0.005 & 0.020 & 0.008 & 0.052 & 0.051 & 0.053 & 0.049 & 0.061 & 0.019 & 0.028 & 0.008 \\
   \hline
  && \multicolumn{8}{c}{\bf (i)} & \multicolumn{8}{c}{\bf (ii)} \\
\cmidrule(lr){3-10} \cmidrule(lr){11-18}
  $n$ & $p$ & dCov & SC & $\text{LD}_\tau$ & $\text{LD}_\rho$ & $\text{LD}_{t^*}$ & CJ & $\text{HL}_\tau$ & $\text{HL}_\rho$  & dCov & SC & $\text{LD}_\tau$ & $\text{LD}_\rho$& $\text{LD}_{t^*}$  & CJ & $\text{HL}_\tau$ & $\text{HL}_\rho$ \\
  \hline
  60 &   50 & 0.058 & 0.146 & 0.067 & 0.057  & 0.064 & 0.974 & 0.027 & 0.016 & 0.055 & 0.073 & 0.066 & 0.056 & 0.086 & 0.377 & 0.033 & 0.017 \\
      60 &  100 & 0.052 & 0.148 & 0.062 & 0.054 & 0.064 & 1.000 & 0.027 & 0.010 & 0.057 & 0.075 & 0.062 & 0.052 & 0.075 & 0.628 & 0.022 & 0.011 \\
      60 &  200 & 0.052 & 0.150 & 0.057 & 0.047 & 0.066 & 1.000 & 0.022 & 0.007 & 0.058 & 0.067 & 0.061 & 0.053 & 0.070 & 0.888 & 0.026 & 0.009 \\
      60 &  400 & 0.057 & 0.147 & 0.060 & 0.049 & 0.075 & 1.000 & 0.021 & 0.004  & 0.057 & 0.073 & 0.063 & 0.053 & 0.074 & 0.992 & 0.020 & 0.004 \\
      60 &  800 & 0.059 & 0.148 & 0.065 & 0.056 & 0.074 & 1.000 & 0.016 & 0.002 & 0.057 & 0.071 & 0.058 & 0.048 & 0.066 & 1.000 & 0.014 & 0.003 \\
     100 &   50 & 0.054 & 0.143 & 0.057 & 0.052 & 0.066 & 0.975 & 0.036 & 0.027 & 0.059 & 0.076 & 0.060 & 0.057 & 0.070 & 0.483 & 0.037 & 0.029 \\
     100 &  100 & 0.056 & 0.154 & 0.056 & 0.050 & 0.063 & 1.000 & 0.034 & 0.023 & 0.060 & 0.075 & 0.059 & 0.052 & 0.064 & 0.774 & 0.034 & 0.020 \\
    100 &  200 & 0.055 & 0.160 & 0.054 & 0.048 & 0.058 & 1.000 & 0.031 & 0.016 & 0.047 & 0.066 & 0.051 & 0.046 & 0.054 & 0.978 & 0.033 & 0.018 \\
    100 &  400 & 0.052 & 0.145 & 0.054 & 0.049 & 0.056 & 1.000 & 0.025 & 0.010 & 0.054 & 0.070 & 0.051 & 0.045 & 0.053 & 1.000 & 0.030 & 0.013 \\
    100 &  800 & 0.053 & 0.142 & 0.060 & 0.056 & 0.053 & 1.000 & 0.020 & 0.010 & 0.054 & 0.067 & 0.057 & 0.050 & 0.064 & 1.000 & 0.023 & 0.010 \\
   \hline
\end{tabular} %
}
\end{table}

\begin{example}
\label{eg:power_normal}
The data $W=(W_1,...,W_p) \in {\mathbb R}^p$ are generated from multivariate normal distribution with the following three covariance matrices $\boldsymbol \Sigma=(\sigma_{ij}(\rho))_{i,j=1}^p$ for $\rho=0.25$.
\begin{itemize}
\item {\bf AR(1) structure:} $\sigma_{ii}=1$ and $\sigma_{ij}=\rho^{|i-j|}$ for all $i,j \in \{1,...,d\}$;
\item {\bf Band structure:} $\sigma_{ii}=1$ for $i=1,...,d$; $\sigma_{ij}=\rho$ if $0 < |i-j| < 3$ and $\sigma_{ij}=0$ if $|i-j| \ge 3$;
\item {\bf Block structure:} Define $\boldsymbol \Sigma_{block}=(\sigma^*_{ij})$ with  $\sigma_{ii}=1$ and $\sigma_{ij}=\rho$ if $i \ne j$ for all $i, j  \in \{1,...,5\}$. The covariance matrix is given by the following Kronecker product $\boldsymbol \Sigma = \boldsymbol I_{\lfloor p/5 \rfloor} \otimes \boldsymbol\Sigma_{block} $.
\end{itemize}
\end{example}

Table \ref{table:power_normal} reports the power from Example \ref{eg:power_normal}. It shows that the ${\mathcal L}_2$ type tests perform equally well with power one for most of the cases, while the maximum type tests endure severe power loss when sample size is small or dimension is high. The reason lies in the fact that the alternatives we consider here are dense and therefore favor the ${\mathcal L}_2$ type tests, whereas the ${\mathcal L}_\infty$ type tests target sparse alternative instead and do not work very well in this case.

\begin{table}[ht]
\centering
\scriptsize
\caption{Power of the tests from Example \ref{eg:power_normal}}
\label{table:power_normal}
\begin{tabular}{ccccccccccc}
  \hline
case & $n$ & $p$ & dCov & SC & $\text{LD}_\tau$ & $\text{LD}_\rho$ &  $\text{LD}_{t^*}$ & CJ & $\text{HL}_\tau$ & $\text{HL}_\rho$ \\
  \hline
\multirow{10}{*}{\bf AR(1)} &   60 &   50 & 0.886 & 0.957 & 0.939 & 0.925 & 0.931 & 0.271 & 0.318 & 0.223 \\
   &   60 &  100 & 0.906 & 0.969 & 0.949 & 0.939 & 0.958 & 0.158 & 0.240 & 0.137 \\
   &   60 &  200 & 0.909 & 0.973 & 0.955 & 0.944 & 0.977 & 0.081 & 0.177 & 0.070 \\
   &   60 &  400 & 0.909 & 0.973 & 0.957 & 0.949 & 0.981& 0.029 & 0.105 & 0.031 \\
   &   60 &  800 & 0.908 & 0.972 & 0.955 & 0.947 & 0.987 & 0.010 & 0.070 & 0.012 \\
   &  100 &   50 & 0.998 & 1.000 & 1.000 & 1.000 & 0.999 & 0.849 & 0.827 & 0.764 \\
   &  100 &  100 & 0.999 & 1.000 & 1.000 & 1.000 &1.000 & 0.795 & 0.790 & 0.698 \\
   &  100 &  200 & 1.000 & 1.000 & 1.000 & 1.000 &1.000 & 0.705 & 0.727 & 0.594 \\
   &  100 &  400 & 0.999 & 1.000 & 1.000 & 1.000 &1.000 & 0.579 & 0.653 & 0.477 \\
   &  100 &  800 & 0.999 & 1.000 & 1.000 & 1.000 &1.000 & 0.428 & 0.573 & 0.353 \\
   \hline
\multirow{10}{*}{\bf Band}   &   60 &   50 & 1.000 & 1.000 & 1.000 & 1.000 &1.000 & 0.427 & 0.494 & 0.368 \\
   &   60 &  100 & 0.999 & 1.000 & 1.000 & 1.000 &1.000 & 0.285 & 0.406 & 0.247 \\
   &   60 &  200 & 1.000 & 1.000 & 1.000 & 1.000 &1.000 & 0.156 & 0.303 & 0.132 \\
   &   60 &  400 & 1.000 & 1.000 & 1.000 & 1.000 &1.000 & 0.065 & 0.196 & 0.056 \\
   &   60 &  800 & 1.000 & 1.000 & 1.000 & 1.000 &1.000 & 0.024 & 0.133 & 0.026 \\
   &  100 &   50 & 1.000 & 1.000 & 1.000 & 1.000 &1.000 & 0.965 & 0.957 & 0.928 \\
   &  100 &  100 & 1.000 & 1.000 & 1.000 & 1.000 &1.000 & 0.946 & 0.943 & 0.894 \\
   &  100 &  200 & 1.000 & 1.000 & 1.000 & 1.000 &1.000 & 0.905 & 0.927 & 0.831 \\
   &  100 &  400 & 1.000 & 1.000 & 1.000 & 1.000 &1.000 & 0.811 & 0.883 & 0.729 \\
   &  100 &  800 & 1.000 & 1.000 & 1.000 & 1.000 & 1.000& 0.668 & 0.807 & 0.578 \\
   \hline
\multirow{10}{*}{\bf Block}   &   60 &   50 & 0.999 & 1.000 & 1.000 & 1.000 &0.999 & 0.442 & 0.503 & 0.372 \\
   &   60 &  100 & 1.000 & 1.000 & 1.000 & 1.000 &1.000 & 0.282 & 0.400 & 0.239 \\
   &   60 &  200 & 1.000 & 1.000 & 1.000 & 1.000 &1.000 & 0.149 & 0.303 & 0.128 \\
   &   60 &  400 & 1.000 & 1.000 & 1.000 & 1.000 &1.000 & 0.065 & 0.191 & 0.058 \\
   &   60 &  800 & 1.000 & 1.000 & 1.000 & 1.000 &1.000 & 0.020 & 0.127 & 0.022 \\
   &  100 &   50 & 1.000 & 1.000 & 1.000 & 1.000 &1.000 & 0.959 & 0.952 & 0.918 \\
   &  100 &  100 & 1.000 & 1.000 & 1.000 & 1.000 &1.000 & 0.936 & 0.935 & 0.880 \\
   &  100 &  200 & 1.000 & 1.000 & 1.000 & 1.000 &1.000 & 0.899 & 0.919 & 0.830 \\
   &  100 &  400 & 1.000 & 1.000 & 1.000 & 1.000 &1.000 & 0.812 & 0.883 & 0.733 \\
   &  100 &  800 & 1.000 & 1.000 & 1.000 & 1.000 &1.000 & 0.666 & 0.805 & 0.571 \\
   \hline
\end{tabular}
\end{table}

\begin{example}
\label{eg:power_mix}
Let $\omega$ be generated from a standard Gaussian distribution with $\omega \sim N_{p/5}(0, I_{p/5})$. The dependence structure is constructed through the non-linear functions such that $W=(g_1(\omega),~g_2(\omega),~g_3(\omega),~g_4(\omega),~g_5(\omega)) \in \mathbb{R}^p $, where $g_1(x)=x, ~g_2(x)=\sin(2\pi x),~ g_3(x)=\cos(2\pi x),~ g_4(x)=\sin(4\pi x)$ and $g_5(x)=\cos(4\pi x)$ and $g_i(\omega)$ means applying the function $g_i$ to each component of $\omega$.
\end{example}

\begin{example}
\label{eg:power_log}
Let $\omega$ be generated from a standard Gaussian distribution with $\omega \sim N_{p/2}(0, I_{p/2})$. The dependence structure is constructed through the non-linear functions such that $W=(g_1(\omega),~g_2(\omega)) \in \mathbb{R}^p $, where $g_1(x)=x$ and $g_2(x)=\log(x^2)$ and $g_i(\omega)$ means applying the function $g_i$ to each component of $\omega$.
\end{example}

\begin{example}
\label{eg:power_sin}
Let $\omega$ be generated from univariate standard normal distribution. The dependence structure is constructed through the non-linear functions such that $W=(\sin(\pi \omega),$  $\sin(2\pi \omega),~...,~\sin(p\pi \omega))$.
\end{example}

Examples \ref{eg:power_mix}, \ref{eg:power_log} and \ref{eg:power_sin} are designed for the non-linear and non-monotone dependence, in which case our dCov-based test as well as the $\text{LD}_{t^*}$ demonstrate the highest power among all the competing methods as seen from Table \ref{table:power_nonmonotone}. However, notice that the power for the proposed test increases as the dimension increases while the $\text{LD}_{t^*}$ shows the opposite pattern. In Section 2 of supplementary material, we presented further comparison between the two tests, where we found our proposed test outperforms the $\text{LD}_{t^*}$ under some non-Gaussian data generating processes, especially when the sample size is small and dimension is low. The other three ${\mathcal L}_2$ type tests only exhibit power in Example \ref{eg:power_sin} and the powers diminish substantially and even down to nominal level in other cases. On the other hand, for the ${\mathcal L}_\infty$ type tests, only $\text{HL}_\tau$ has some power in detecting the non-monotone dependence; the other two maximum type tests maintain the power around nominal level $\alpha$. These examples clearly demonstrate the advantage of the distance covariance based test in identifying the non-linear and non-monotone dependence among the data.

\begin{table}[ht]
\centering
\scriptsize
\caption{Power performance for detecting non-monotone dependence}
\label{table:power_nonmonotone}
\begin{tabular}{ccccccccccc}
  \hline
 & $n$ & $p$ & dCov & SC & $\text{LD}_\tau$ & $\text{LD}_\rho$ &  $\text{LD}_{t^*}$ & CJ & $\text{HL}_\tau$ & $\text{HL}_\rho$ \\
  \hline
\multirow{10}{*}{\bf Example \ref{eg:power_mix}} &   60 &   50 & 1.000 & 0.037 & 0.127 & 0.055 &1.000 & 0.022 & 0.261 & 0.044 \\
   &   60 &  100 & 1.000 & 0.038 & 0.121 & 0.057 &1.000& 0.014 & 0.299 & 0.032 \\
   &   60 &  200 & 1.000 & 0.039 & 0.126 & 0.059 &1.000& 0.009 & 0.332 & 0.022 \\
   &   60 &  400 & 1.000 & 0.033 & 0.117 & 0.054 &1.000& 0.006 & 0.369 & 0.017 \\
   &   60 &  800 & 1.000 & 0.033 & 0.114 & 0.057 &1.000& 0.004 & 0.403 & 0.011 \\
   &  100 &   50 & 1.000 & 0.036 & 0.123 & 0.049 &1.000& 0.032 & 0.285 & 0.059 \\
   &  100 &  100 & 1.000 & 0.037 & 0.116 & 0.055 &1.000& 0.028 & 0.337 & 0.054 \\
   &  100 &  200 & 1.000 & 0.036 & 0.117 & 0.056 &1.000& 0.025 & 0.390 & 0.046 \\
   &  100 &  400 & 1.000 & 0.035 & 0.114 & 0.051 &1.000& 0.016 & 0.424 & 0.033 \\
   &  100 &  800 & 1.000 & 0.037 & 0.115 & 0.054 &1.000& 0.013 & 0.464 & 0.025 \\
   \hline
\multirow{10}{*}{\bf Example \ref{eg:power_log}}   &   60 &   50 & 1.000 & 0.054 & 0.257 & 0.109 &1.000& 0.035 & 0.302 & 0.050 \\
   &   60 &  100 & 1.000 & 0.054 & 0.266 & 0.109 &1.000& 0.030 & 0.336 & 0.033 \\
   &   60 &  200 & 1.000 & 0.052 & 0.260 & 0.111 &1.000& 0.039 & 0.378 & 0.028 \\
   &   60 &  400 & 1.000 & 0.059 & 0.271 & 0.112 &1.000& 0.031 & 0.420 & 0.016 \\
   &   60 &  800 & 1.000 & 0.055 & 0.261 & 0.104 &1.000& 0.032 & 0.476 & 0.011 \\
   &  100 &   50 & 1.000 & 0.049 & 0.264 & 0.109 &1.000& 0.046 & 0.334 & 0.062 \\
   &  100 &  100 & 1.000 & 0.057 & 0.259 & 0.114 &1.000& 0.046 & 0.384 & 0.059 \\
   &  100 &  200 & 1.000 & 0.048 & 0.253 & 0.106 &1.000& 0.061 & 0.436 & 0.048 \\
   &  100 &  400 & 1.000 & 0.051 & 0.257 & 0.104 &1.000& 0.066 & 0.486 & 0.038 \\
   &  100 &  800 & 1.000 & 0.052 & 0.252 & 0.107 &1.000& 0.083 & 0.535 & 0.030 \\
   \hline

\multirow{10}{*}{\bf Example \ref{eg:power_sin}}    &   60 &   50 & 0.694 & 0.609 & 0.607 & 0.591 &1.000& 0.020 & 0.201 & 0.028 \\
   &   60 &  100 & 0.769 & 0.728 & 0.718 & 0.706 &0.978& 0.015 & 0.200 & 0.018 \\
   &   60 &  200 & 0.828 & 0.807 & 0.808 & 0.797 &0.923& 0.013 & 0.203 & 0.012 \\
   &   60 &  400 & 0.887 & 0.873 & 0.874 & 0.867 &0.817& 0.008 & 0.193 & 0.008 \\
   &   60 &  800 & 0.919 & 0.904 & 0.896 & 0.898 &0.703& 0.004 & 0.183 & 0.003 \\
   &  100 &   50 & 0.771 & 0.609 & 0.617 & 0.593 &1.000& 0.027 & 0.390 & 0.067 \\
   &  100 &  100 & 0.800 & 0.732 & 0.725 & 0.716 &1.000& 0.023 & 0.411 & 0.053 \\
   &  100 &  200 & 0.843 & 0.808 & 0.805 & 0.800 &1.000& 0.021 & 0.450 & 0.042 \\
   &  100 &  400 & 0.887 & 0.857 & 0.859 & 0.857 &1.000& 0.015 & 0.484 & 0.028 \\
   &  100 &  800 & 0.918 & 0.902 & 0.901 & 0.898 &0.989& 0.011 & 0.502 & 0.020 \\
  \hline
\end{tabular}
\end{table}

\subsection{Tests for joint dependence}
\label{sec:buhlmann}
As mentioned in the introduction, our test mainly focuses on the presence of the ``main effects'' of joint dependence and tests for the sub-null $H_0'$. In comparison, dHSIC proposed by \cite{Pfister:2016aa} targets at the joint (mutual) dependence.  As discussed in Section \ref{sec:joint}, the theory for dHSIC is restricted to the fixed dimensional case and its validity in the high dimensional case is unknown. Here we compare our proposed method with dHSIC and MdCov under different scenarios.

Since dHSIC test and MdCov require that $n \ge 2p$,  we choose three combinations $n=60, p=18$; $n=100, p=36$ and $n=200, p=72$. We compare the three tests for some of the examples chosen from  Section \ref{sec:sim_mutual}, namely Example \ref{eg:size}, \ref{eg:power_normal}, \ref{eg:power_log} and \ref{eg:power_sin}. Besides, we also consider an interesting example as follows, where $W$ is pairwise independent but not jointly independent.
\begin{example}
\label{eg:mutual}
Consider the tuple of three random variables $\boldsymbol{Z} = (Z_1, Z_2, Z_3)$, where $Z_1$, $Z_2$ are independent Bernoulli random variables with success probability $1/2$, $Z_3 = \boldsymbol{1}_{(Z_1=Z_2)}$ and $\boldsymbol{1}_{(\cdot)}$ is the indicator function. Our data consists of $p/3$ i.i.d copies of $\boldsymbol{Z}$, that is, $W=(\boldsymbol{Z}_1,\dots,\boldsymbol{Z}_{p/3})$.
\end{example}

The size and power (rejection probabilities) are reported based on 5000 Monte Carlo simulations at the nominal level $\alpha=0.05$. Here the dHSIC and MdCov (with $a=1$) are implemented as permutation tests; we use Gaussian kernel for dHSIC where the bandwidth parameter $\gamma$ is chosen as $\gamma_0=$ the median of all pairwise distances [see \cite{gretton2012}].
Following the suggestion of an anonymous reviewer, we also examine the sensitivity of dHSIC with respect to the choice of $\gamma$ by letting $\gamma=c\gamma_0$, with $c=\frac{1}{9}, \frac{1}{3},3,9$ (denoted as dHSIC(c) in Table~\ref{table:buhlmann}).

Table \ref{table:buhlmann} summarizes the rejection rates for the above mentioned three tests. We note that  dHSIC delivers zero rejection rates for all cases in Example \ref{eg:size} when $(n,p)=(200,72)$ and  $c=1/9,1/3,1,3$, and when  $(n,p)=(100,36)$ and $c=c(1/9,1/3)$. A careful look at the source code from ``dHSIC" package in CRAN indicates that when the dimension is too high the sample statistic as well as the ones based on the permuted samples become a constant when the underlying data is generated on the real line. This results in a zero rejection rate.
This suggests that the smaller the bandwidth is, the more limited range of dimensionality dHSIC can handle.
For $(n,p)=(60,18)$, the performance of dHSIC with various bandwidth seems more reasonable and we shall comment on that below.
 When the data are jointly independent as in Example \ref{eg:size}, all tests have quite accurate rejection rates around the nominal level 5\%, which suggests that normal approximation works quite well for our test even when $p=18$ and $n=60$.
 For linearly dependent and non-linearly dependent data in Example \ref{eg:power_normal} and Examples \ref{eg:power_log}-\ref{eg:power_sin} respectively, dCov demonstrates consistently high power against the null than both MdCov and dHSIC$(c)$ for all $c$; surprisingly, dHSIC almost has no power for linearly dependent data in Example \ref{eg:power_normal} except when $c=9$. In both Example \ref{eg:power_normal} and Example \ref{eg:power_log}, we see the power monotonically increases with respect to $c$, suggesting larger bandwidth brings more power for these examples, although the power with dHSIC(9) decreases when $(n,p)$ increases; MdCov has no power or very little power for the dependent data. For Example \ref{eg:mutual}, the data is pairwise independent but not jointly independent, thus our test cannot detect any dependence beyond the pairwise dependence and has rejection rate around the nominal level, which is consistent with our expectation; dHSIC has a reasonable rejection rate when dimension is small relative to the sample size, but endures severe power loss when the dimension is high. The fact that the power for dHSIC is so low when $(n,p)=(200,72)$ is somewhat expected, since most of triples in Example \ref{eg:mutual} are mutually independent (as mentioned in \cite{sun1998}) and thus the data with dimension $p=72$ are less mutually dependent than that when $p=36$ and $18$.  Additionally, within dHSIC based tests, the choice of $\gamma=3\gamma_0$ corresponds to the highest power, suggesting that larger bandwidth does not always bring more power and the optimal bandwidth depends on the data generating process.
 These findings suggest that (i) incapability of the dHSIC/MdCov to handle high dimensional data is quite apparent. Taking larger bandwidth in dHSIC may help to alleviate the impact of  high dimensionality, but the results in Examples \ref{eg:power_normal}, \ref{eg:power_log}, \ref{eg:power_sin} and \ref{eg:mutual}  indicate that there might be intrinsic difficulty to capture all kinds of higher order dependence beyond pairwise dependence when the dimension is high; (ii) the usual rule of thumb choice $\gamma_0$ for the bandwidth parameter of Gaussian kernel in dHSIC (or HSIC) works well in the low dimensional setting, but the performance in high dimension is sub-optimal, and taking a larger bandwidth could improve the power substantially in some examples. How to choose a
 good bandwidth parameter remains an important open problem for dHSIC.

% Overall, our dCov test did a reasonable job in testing the mutual independence without any tuning parameters like the kernel and hyper-parameters  choices for dHSIC; we also enjoy the constraint-free setting for sample size $n$ and dimension $p$ in our test.

\begin{table}[h!]
\scriptsize
\centering
\caption{Comparison between proposed dCov test and dHSIC test in Section \ref{sec:buhlmann}}
\label{table:buhlmann}
\begin{tabular}{cccccccccccc}
  \hline
  \hline
    &   &  & \multicolumn{4}{c}{Example \ref{eg:size}} & \multicolumn{2}{c}{Example \ref{eg:power_normal}} & \multicolumn{3}{c}{Example} \\
\cmidrule(lr){4-7}
\cmidrule(lr){8-9}
\cmidrule(lr){10-12}
  &   n & p & (i) & (ii) & (iii) & (iv) & AR & Band & \ref{eg:power_log} & \ref{eg:power_sin} & \ref{eg:mutual}  \\
  \hline
{\multirow{3}{*}{dCov}}  & 60 & 18 & 0.051 & 0.055 & 0.056 & 0.052 & 0.782 & 0.990 & 1.000 & 1.000 & 0.051\\
& 100 & 36 & 0.051 & 0.051 & 0.048 & 0.049 & 0.995 & 1.000 & 1.000 & 1.000 & 0.048\\
& 200 & 72 & 0.055 & 0.049 & 0.049 & 0.052 & 1.000 & 1.000 & 1.000 & 1.000 & 0.057\\
\hline
{\multirow{3}{*}{MdCov}}  & 60 & 18 & 0.059 & 0.052& 0.058 &0.056 &0.089 &0.124& 0.000 &0.000 &0.030 \\
& 100 & 36 & 0.054 & 0.058 & 0.053 & 0.054 &0.088& 0.121 & 0.000 &0.000 &0.015\\
& 200 & 72 & 0.051 & 0.050 & 0.054 & 0.058 & 0.082& 0.123 &0.000 &0.000 &0.010\\
   \hline
 {\multirow{3}{*}{dHSIC(1/9)}}  & 60 & 18  &  0.041 & 0.057 & 0.052 & 0.044 & 0.021 & 0.010 & 0.071 & 1.000 & 0.387 \\
  & 100 & 36 &   0.000 & 0.000 & 0.000 & 0.000 & 0.000 & 0.000 & 0.000 & 1.000 & 0.060 \\
   & 200 & 72 &  0.000 & 0.000 & 0.000 & 0.000 & 0.000 & 0.000 & 0.000 & 1.000 & 0.000 \\
    \hline
 {\multirow{3}{*}{dHSIC(1/3)}}  & 60 & 18 &   0.053 & 0.054 & 0.052 & 0.047 & 0.025 & 0.010 & 0.136 & 1.000 & 0.341 \\
  & 100 & 36 &   0.000 & 0.000 & 0.000 & 0.000 & 0.000 & 0.000 & 0.000 & 1.000 & 0.064 \\
   & 200 & 72 &  0.000 & 0.000 & 0.000 & 0.000 & 0.000 & 0.000 & 0.000 & 1.000 & 0.000 \\
    \hline
{\multirow{3}{*}{dHSIC(1)}}  & 60 & 18 & 0.050 & 0.052 & 0.049 & 0.048 & 0.070 & 0.091 & 0.267 & 1.000 & 0.664\\
& 100 & 36 & 0.050 & 0.045 & 0.044 & 0.051 & 0.036 & 0.036 & 0.173 & 1.000 & 0.282\\
& 200 & 72 & 0.000 & 0.000 & 0.000 & 0.000 & 0.000 & 0.000 & 0.000 & 1.000 & 0.072\\
   \hline
 {\multirow{3}{*}{dHSIC(3)}}  & 60 & 18 &  0.050 & 0.051 & 0.047 & 0.046 & 0.078 & 0.115 & 0.340 & 1.000 & 0.708 \\
  & 100 & 36 &    0.053 & 0.045 & 0.046 & 0.045 & 0.077 & 0.118 & 0.198 & 1.000 & 0.314 \\
& 200 & 72 &   0.000 & 0.000 & 0.000 & 0.000 & 0.000 & 0.000 & 0.000 & 1.000 & 0.073 \\
    \hline
 {\multirow{3}{*}{dHSIC(9)}}  & 60 & 18 &   0.049 & 0.050 & 0.051 & 0.050 & 0.297 & 0.619 & 0.775 & 1.000 & 0.389 \\
  & 100 & 36 &   0.051 & 0.045 & 0.050 & 0.047 & 0.216 & 0.507 & 0.432 & 1.000 & 0.292 \\
  & 200 & 72 &    0.049 & 0.051 & 0.025 & 0.051 & 0.142 & 0.324 & 0.168 & 1.000 & 0.072 \\
    \hline
    \hline
\end{tabular}
\end{table}

% % % % % % % % % % % % % %
% % % % % % % % % % % % % %

\section{Data Illustration}
\label{sec:data}
In this section, we employ the proposed methods to analyze the prostate cancer data set and report the results. The original prostate cancer data was analyzed by \cite{adam2002} to study the protein profiling technologies that can simultaneously resolve and analyze multiple proteins in early detection of prostate cancer. Surface enhanced laser desorption/ionization mass spectrometry protein profiles of patients' blood serum samples are recorded. These profiles contains the intensity values for a large amount of time-of-flight values. The time-of-flight is related to the mass over charge ratio $m/z$ of the constituent proteins in the blood. There are 157 healthy patients and 167 prostate cancer patients with 48,538 $m/z$-sites in total.

This data set has been analyzed by several statisticians for various purposes. Following previous researchers, the $m/z$-sites below 2000 are ignored due to the possible chemical artifacts occurrence under that level. \cite{tibshirani2005} averaged the intensity values in consecutive blocks of 20, which gives a total of 2181 dimensions per serum sample. \cite{levina2008}, \cite{qiu2012} further averaged the data of \cite{tibshirani2005}  in consecutive blocks of 10, resulting in a total of 218 dimensions. We follow this approach and consider the observation $\boldsymbol{w}_i = (w_{i,1}, . . . , w_{i,218})$ with intensity profile of length 218 for patient $i$ to test the mutual independence and the banded dependence structure if the former hypothesis is rejected.

\begin{figure}[t!]
\centering
\label{fig:prostate}
\includegraphics[width=5in]{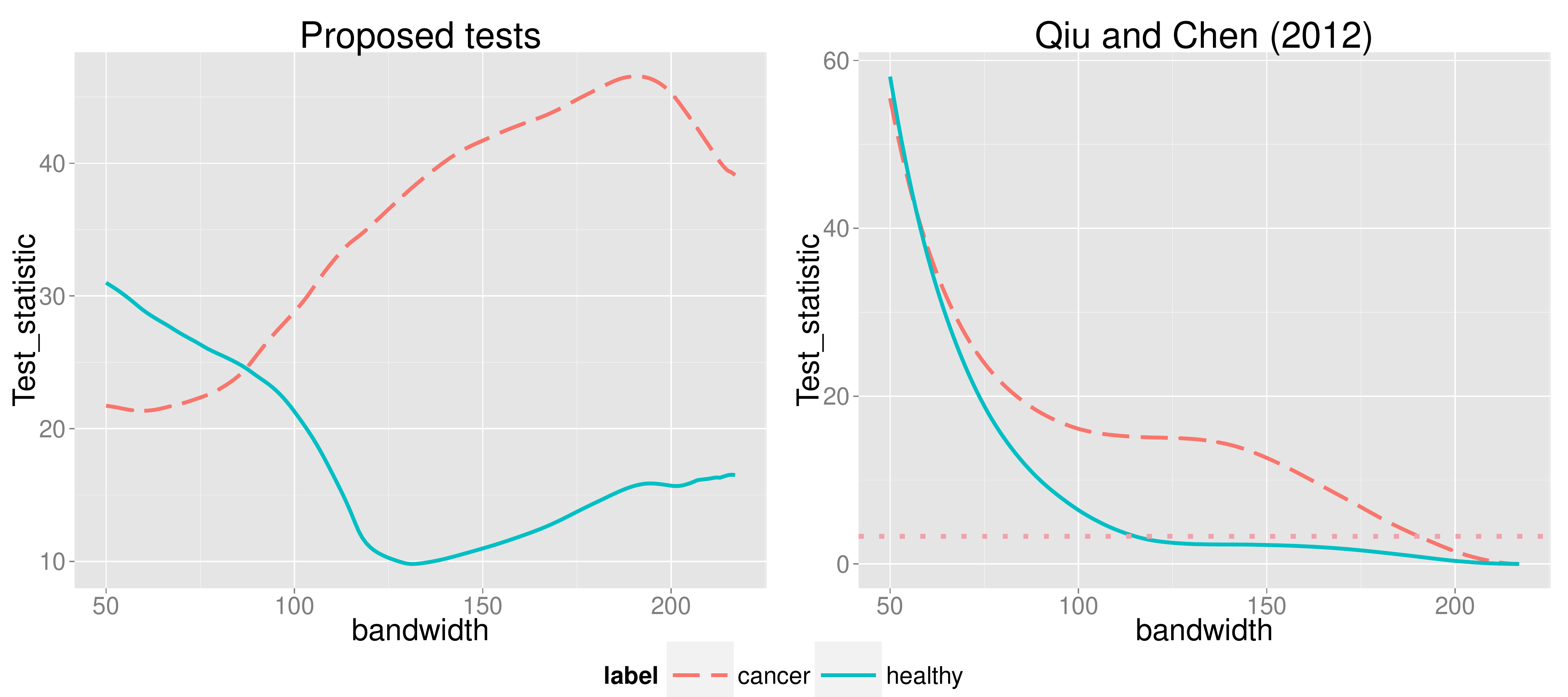}
\caption{Values of test statistics for healthy and prostate cancer patients of the proposed test (left panel) and \cite{qiu2012} test (right panel).}
\end{figure}

We conduct the analysis for two groups separately: the healthy group (157 samples), prostate cancer group (167 samples). The tests for mutual independence are both rejected for these two groups with p-values 0. Since there is a natural ordering for these 218 dimensions ($m/z$-sites), we further carry out the banded dependence structure test with given bandwidth $h$ from 50 to 217. The corresponding values of the test statistics are plotted in Figure \ref{fig:prostate}. We also employ the proposed methods to the mixed group data (157 healthy patients together with 167 prostate cancer patients), but the results are not informative and therefore omitted. Some previous studies also used the standardized data and we found no significant differences between using the original data and the standardized data in our tests for this particular prostate cancer data set.

The test results suggest that the dependence structure is not banded for both the patient group and healthy group. The shape of the curve from healthy group in the left panel of Figure \ref{fig:prostate} indicates that the overall dependence is decreasing steeply first and then increasing gradually as the bandwidth increases; moreover, the test statistics are the smallest for $125 \le h \le 150$, which hints at that the dependence is relatively weak for those bandwidths. The curve from prostate cancer group, however, demonstrates completely different pattern. It increases substantially from $h=70$ to $h=185$ and then decrease afterwards, which suggests strong non-linear dependence. The sharp contrast between healthy group and cancer group signifies significant differences in the dependence structure for prostate cancer and non-cancer people.

\cite{bickel2008}, \cite{qiu2012} (the test statistic values are shown in the right panel of Figure \ref{fig:prostate}) used covariance matrix based method and concluded that the healthy group's covariance matrix is likely to be banded with bandwidth 144 and 121 respectively and may not be banded at all for the prostate cancer group. Our method implies that the dependence structure is not banded for both groups and the non-linear dependence is especially strong between bandwidth 90 and 185 for the cancer group.

%\subsection{Colon data and Leukemia data}

%We next apply our method to two widely used datasets: the colon data of Alon et al. (1999) and the leukemia data of Golub et al. (1999). The colon data contains gene expression in 40 tumors and 22 normal colon tissue samples using Affymetrix oligonucleotide array complementary to more than 6,500 human genes and only the top 2000 genes with the highest minimal intensity across the samples were kept. The leukemia data includes 3571 gene expression levels of patients suffering from either acute myeloid leukemia (AML) or acute lymphoblastic leukemia (ALL), with 47 and 25 patients in each category respectively.

%The test results suggest that the gene expression are not mutually independent for both of the data sets. Specifically, the test statistic for 62 samples in colon data is 4028.01, and 3083.01 for tumor tissue, 1440.56 for normal tissue; as for leukemia data the test statistic is 3034.33642 for all 72 patients and 909.55 for AML group 1926.82 for ALL group. The results also imply that the dependence is stronger in tumor tissue and acute lymphoblastic leukemia patients than normal tissue and acute myeloid leukemia respectively.

\section{Conclusion}
\label{sec:conclusion}

 In the present paper, we proposed a mutual independence test using sum of pairwise squared distance covariance and further extended the test to testing the banded dependence structure. Asymptotic distributions of the test statistics were studied under the null and local alternatives using tools related to U-statistics. We view our new test as a useful addition to the family of mutual independence tests, for example, \cite{schott2005}, \cite{cai2011}, \cite{han2014}, \cite{leung2015} among others, as few existing tests can capture non-monotonic dependence. Our numerical results demonstrate the merit of the proposed test in identifying the non-linear and non-monotonic dependence in the data compared with Pearson correlation and rank correlation based counterparts, which only focus on linear dependence and monotone dependence respectively. Compared to \cite{bergsma2014}'s $t^*$-based test, our test is more computationally efficient, has less size inflation and comparable power in all examples examined.

 As mentioned early, sum of squares/${\cal L}^2$ type statistic naturally targets at non-sparse but weak alternatives. It would be interesting to consider the ${\cal L}_{\infty}$/maximum type statistic using the distance covariance in the future to capture sparse and strong dependence. The mild size distortion for our test at small sample size may be alleviated by using permutation-based critical values. However, permutation based test becomes quite expensive in high dimension, and it will be interesting to develop more accurate approximation of our null distribution with manageable/scalable computational cost. Furthermore, we can use distance correlation based test in testing mutual independence or consider a more general multivariate dependence measure instead of pairwise dependence measure to capture the dependence of any three or more subsets of $p$ components, which is certainly more challenging and is left for future work.

\section{Acknowledgement}
Zhang acknowledges partial financial support from NSF grant DMS-1607320 and Shao acknowledges partial financial support from NSF grants DMS-1407037 and DMS-1607489. We would like to thank the two reviewers, associated editor and the co-editor Piotr Fryzlewicz for their
constructive comments that led to a substantial improvement of the article.

\begin{center} SUPPLEMENTARY MATERIAL
\end{center}
The supplementary material contains all the technical details of the main theoretical results and some additional numerical comparison.

\appendix

\bibliographystyle{agsm}
\footnotesize
\bibliography{reference}

\end{document}